\documentclass[]{aa}
\usepackage{graphicx}
\usepackage{supertabular}
\usepackage{txfonts}
\newcommand{\kms}{\,km\,s$^{-1}$}

\newcommand{\kup}{KU\,Peg}
\newcommand{\kupt}{KU\,Peg }

\begin{document}

   \title{Time-series Doppler images and surface differential rotation of the effectively-single
rapidly-rotating K-giant KU\,Pegasi\thanks{Based on data obtained with the STELLA robotic observatory in Tenerife, an AIP facility jointly operated by AIP and IAC.}}

\author{Zs.~K\H{o}v\'ari\inst{1}
\and A.~K\"unstler\inst{2}
\and K.~G. Strassmeier\inst{2}
\and T.~A. Carroll\inst{2}
\and M.~Weber\inst{2}
\and L.~Kriskovics\inst{1}
\and K.~Ol\'ah\inst{1}
\and K.~Vida\inst{1}
\and T.~Granzer\inst{2}
}

\offprints{Zs. K\H{o}v\'ari}

\institute{Konkoly Observatory,
Research Centre for Astronomy and Earth Sciences, Hungarian Academy of Sciences,
Konkoly Thege \'ut 15-17., H-1121, Budapest, Hungary\\
  \email{kovari@konkoly.hu}
  \and Leibniz Institute for Astrophysics (AIP), An der Sternwarte 16,
D-14482 Potsdam, Germany}

   \date{Received ; accepted}

\abstract  % context heading (optional) leave it empty if necessary
{According to most stellar dynamo theories, differential rotation (DR) plays a crucial role for the generation of toroidal magnetic fields. Numerical models predict surface differential rotation to be anti-solar for rapidly-rotating giant stars, i.e., their surface angular velocity could increase with stellar latitude. However, surface differential rotation has been derived only for a handful of individual giant stars to date. }
{The spotted surface of the K-giant KU\,Pegasi is investigated in order to detect its time evolution and quantify surface differential rotation.}
% and relate it to other (giant) stars.}
{We present altogether 11 Doppler images from spectroscopic data collected with the robotic telescope STELLA between 2006--2011. All maps are obtained with the surface reconstruction code \emph{iMap}. Differential rotation is extracted from these images by detecting systematic (latitude-dependent) spot displacements. We apply a cross-correlation technique to find the best differential rotation law.}
{The surface of \kupt shows cool spots at all latitudes and one persistent warm spot at high latitude. A small cool polar spot exists for most but not all of the epochs. Re-identification of spots in at least two consecutive maps is mostly possible only at mid and high latitudes and thus restricts the differential-rotation determination mainly to these latitudes. Our cross-correlation analysis reveals solar-like differential rotation with a surface shear of $\alpha=+0.040\pm0.006$, i.e., approximately five times weaker than on the Sun. We also derive a more accurate and consistent set of stellar parameters for \kupt including a small Li abundance of ten times less than solar.}
{}


\authorrunning{K\H{o}v\'ari et al.}
\titlerunning{Time-series Doppler images and differential rotation of KU\,Peg}

   \maketitle

%
%________________________________________________________________

\section{Introduction}

Quantifying differential surface rotation has proven difficult even for the Sun. Stellar observations
are even more demanding and correspondingly ambiguous are the results. However, quantitative
detections are now  possible for those stars where we are able to spatially resolve the stellar disk
by means of Doppler imaging. Such observations
\citep[e.g.,][etc.]{2007A&A...476..881K,2007AN....328.1075W,2013A&A...551A...2K,2015A&A...573A..98K}
as well as theoretical considerations \citep{2004AN....325..496K,2012AN....333.1028K} imply that stellar
surface rotation could probably be more complex for evolved stars when compared to main sequence
stars like the Sun. Differential rotation for main-sequence and pre-main sequence stars is found
to decrease with effective temperature \citep{2005MNRAS.357L...1B, 2013A&A...560A...4R} just like
predicted from mean-field dynamo models \citep{2015csss...18..535K}.
However, the situation seems to be less well defined for post-main sequence stars with their much deeper
convective envelopes. Because many of these giants are components in RS\,CVn-type binary systems, their
surfaces are possibly distorted by and respond to the orbital dynamics. Moreover, as a star evolves up the
red-giant branch, its core experiences a modification of nuclear reactions followed by core contraction and
envelope expansion long before helium burning sets in. After the core hydrogen is exhausted,
the hydrogen fusion keeps going in a surrounding shell, providing more helium onto the contracting inert core.
The contraction heats up the core together with the interlocked shell, which expands inward.
At a point the core becomes degenerate. The increasing density at the bottom of the H-rich shell
yields a more efficient H-burning, which eventually blows up the envelope.
The temperature of the envelope decreases and the outer layers become fully convective,
transporting more flux outwards, which explains the rapidly increasing luminosity with decreasing
surface temperature along the RGB.
The shell material penetrates into the hotter regions below, decaying light elements and triggering a mixing process called
the first dredge up, which is responsible for the dilution of the lithium.
Indeed, according to \citet{1994A&A...282..811C,1995ApJ...453L..41C} in low mass ($\le2M_{\odot}$) stars, further rotationally induced mixing occurs after the completion of the first dredge up
\citep[see also][]{1992A&A...265..115Z}.
Such mixing episodes can be inferred from the lowering of the observed surface abundances of the most fragile elements ($^{7}$Li, $^{12}$C) and the $^{12}$C/$^{13}$C isotopic ratio \citep{2000A&A...354..169G}.
%Since the degenerate core material is much more dense than the shell density at the bottom
 Even the simple expansion appears to have an
effect on the mixing of the convective envelope and eventually also alters the surface DR profile as well.
In some cases, the DR profile can be even of anti-solar type, i.e., the equator rotating slower than the
poles  \citep[][etc.]{1991LNP...380..297V,2003A&A...408.1103S,2015A&A...573A..98K}. Whether such anti-solar
DR was already present during the main sequence phase of such a star or explicitly developed during the
expansion phase on the giant branch is not known. Besides, DR of either solar or anti-solar has been derived only for a handful of
 late-type evolved stars to date. A short list of the giant stars with known DR from Doppler imaging would include the following ones. Among the single (or effectively single) giants, solar type DR was reported for FK\,Com \citep{2007A&A...476..881K}, V390\,Aur \citep{2012A&A...541A..44K} and KU\,Peg \citep{2001A&A...373..974W}, i.e., the star to be revisited in this paper, while anti-solar DR was detected on HD\,31993 \citep{2003A&A...408.1103S}, DI\,Psc and DP\,CVn \citep{2013A&A...551A...2K,2014A&A...571A..74K}.
In binary systems solar type DR was found e.g., on the evolved components of
$\zeta$\,And \citep{2012A&A...539A..50K}, XX\,Tri \citep{2015A&A...578A.101K}, and IL\,Hya \citep{2014IAUS..302..379K}, while anti-solar DR was found on $\sigma$\,Gem \citep{2015A&A...573A..98K}, IM\,Peg, UZ\,Lib, etc. \citep[see][and the references therein]{2014SSRv..186..457K}. Our foremost aim is to enlarge the
observational sample of reliable DR detections on giant stars.

Time-series Doppler imaging has proven to be extremely useful for studying stellar DR
\citep[e.g.][]{1996IAUS..176..245V,1997MNRAS.291....1D,1998A&A...330.1029W,2004AN....325..221P}.
When having subsequent Doppler reconstructions of the spotted stellar surface, the rotation rates of
individual spots can reveal the latitude-dependent stellar rotation profile. However, such Doppler
reconstructions require high-resolution spectroscopic time-series data, covering at least two but
better many  consecutive rotation cycles. That this is indeed a challenge for stars with rotation periods
of close to a month is obvious. A unique possibility for such long-term Doppler observations
\citep[see, e.g.,][etc.]{2015A&A...573A..98K,2015A&A...574A..31S,2015A&A...578A.101K} is provided by the
STELLA robotic observatory of the AIP in Tenerife \citep{2010AdAst2010E..19S}. In this paper
we present and analyze such spectroscopic observations of the rapidly-rotating %but long-period
($P_{\rm rot}\approx24$\,days) K-giant \object{KU~Peg} (=HD\,218153).

Chromospheric activity of \kupt was recognized by \citet{1983AJ.....88.1182B} who reported
strong Ca\,{\sc ii}\,H\&K emission. The large chromospheric fluxes were later confirmed with IUE
observations by \citet{1992A&A...254L..36D}. Just recently, \citet{2015A&A...574A..90A}
detected magnetic fields on KU\,Peg and found that the star follows the
magnetic field strength-rotation relationship established for active giants, indicating
that probably a solar type magnetic dynamo was working inside. In addition, the authors
reported an unusually strong X-ray luminosity of $L_X=11.8\times10^{30}$\,erg\,s$^{-1}$
confirming the existence of coronal activity as well.
%yielding $\log\frac{L_X}{L_{\rm bol}}\approx-4.45$ (cf. Table~\ref{T4} in this paper).
KU\,Peg was found to be a single-lined spectroscopic binary with an orbital period of $\approx$1400\,days
\citep{1992A&A...254L..36D}, suggesting that it is \emph{effectively} a single star. Because differential rotation
is supposed to be weakened (or totally quenched) by tidal forces in close binaries
\citep{1981ApJ...246..292S,1982ApJ...253..298S}, \kupt is a good candidate for a comparison with theory.
A projected rotational velocity of 29\,\kms\ was measured by \citet{1997PASP..109..514F}, which
placed the star among the possible Doppler-imaging candidates \citep{2000A&AS..142..275S}.

The first and so far only Doppler-imaging study of \kupt was carried out by \citet[][hereafter Paper~I]{2001A&A...373..974W}
using high-resolution spectra taken with the McMath-Pierce solar telescope and the coud\'e feed telescope at
Kitt Peak National Observatory over two months in 1996/97. The data allowed the reconstruction of two consecutive
Doppler images that revealed an asymmetric polar spot and several other cool spots at lower latitudes. The time evolution
of the spotted surface was followed by means of a cross-correlation analysis and revealed a complex DR profile that
resembled the solar case only in the directional sense, i.e., lower latitudes rotating faster. Its lap time was twice as long as
that of the Sun for the full pole-to-equator range but twice as short if only the latitudes where the Sun has spots was
considered. Moreover, patterns of local meridional flows were detected, which likely play also an important role
for stellar dynamos \citep{2004AN....325..496K,2011AN....332...83K}.

The current paper is organized as follows. In Sect.~\ref{obs} we describe our photometric and spectroscopic
observations. In Sect.~\ref{prot} photometric data from more than 18 years are employed to derive a precise
average rotation period. These data are also used to search for photometric signals of surface DR. In
Sect.~\ref{di}, we first redetermine the basic astrophysical properties of \kupt by including our new photometric
and spectroscopic data. Then we give a brief description of our inversion code \emph{iMap} and its data assumptions
for image reconstruction and, thirdly, we present the time-series Doppler images. In Sect.~\ref{ccf} the consecutive
Doppler images are used to derive the surface DR of \kup. Lithium abundance determination is carried out in Sect.~\ref{lithium}.
The results are summarized and discussed in Sect.~\ref{disc}.

\section{Observations}\label{obs}

\subsection{Photometry}

Photometric observations in this paper were obtained with the Amadeus 0.75\,m automatic photoelectric
telescope (T7-APT) of the AIP operated at Fairborn Observatory in southern Arizona \citep{1997PASP..109..697S}.
The data set consists of altogether 1243 measurements in Johnson $V$ and 1306 measurements in the
Johnson-Cousins $I_C$ band. A total of, so far, 18 years between JD 2\,450\,395 and 2\,457\,015 are covered.
Differential photometric observations were carried out with respect to HD\,218610 as the comparison star
($V=7\fm7940$, $V-I=1\fm34$), and HD\,219050 as the check star. Mean photometric errors were
$0\fm006$ in $V$ and $0\fm009$ in $I_C$. For more details on APT performance and operation,
as well as its data reduction, we refer to \citet{2001AN....322..325G}.

\subsection{Spectroscopy}

A total of 193 high-resolution spectra were collected with the robotic 1.2\,m STELLA-I telescope at the Iza\~{n}a Observatory
in Tenerife, Spain \citep{2010AdAst2010E..19S} during 2006--2011. The STELLA robotic observatory (also containing a 1.2\,m
photometric telescope, now STELLA-I) runs fully autonomous without any personnel on site just guided by weather and meteorological
parameters and the target schedule. The STELLA-II telescope is equipped with the fibre-fed fixed-format
STELLA Echelle Spectrograph (SES). (Note that for most of the time span in the present paper, the SES fibre was connected to one of the
two Nasmyth foci of STELLA-I but was moved to STELLA-II in 2010 after the wide-field imager was inaugurated.) All SES spectra
cover the wavelength range 3900--8800\,\AA\ with a 2-pixel resolution of $R=55\,000$ corresponding to a spectral
resolution of 0.12\,\AA\ at 6500\,\AA. For further details of the performance of the system and also for the detailed
data-reduction procedures, we refer to \citet{2008SPIE.7019E..0LW,2012SPIE.8451E..0KW} and \citet{2011A&A...531A..89W}.
Given in the Appendix, Table~\ref{Tab1}  is a log of all SES observations of \kup.

% ----------------------- F1
%
%
\begin{figure}[b]
\includegraphics[angle=0,width=1.0\columnwidth]{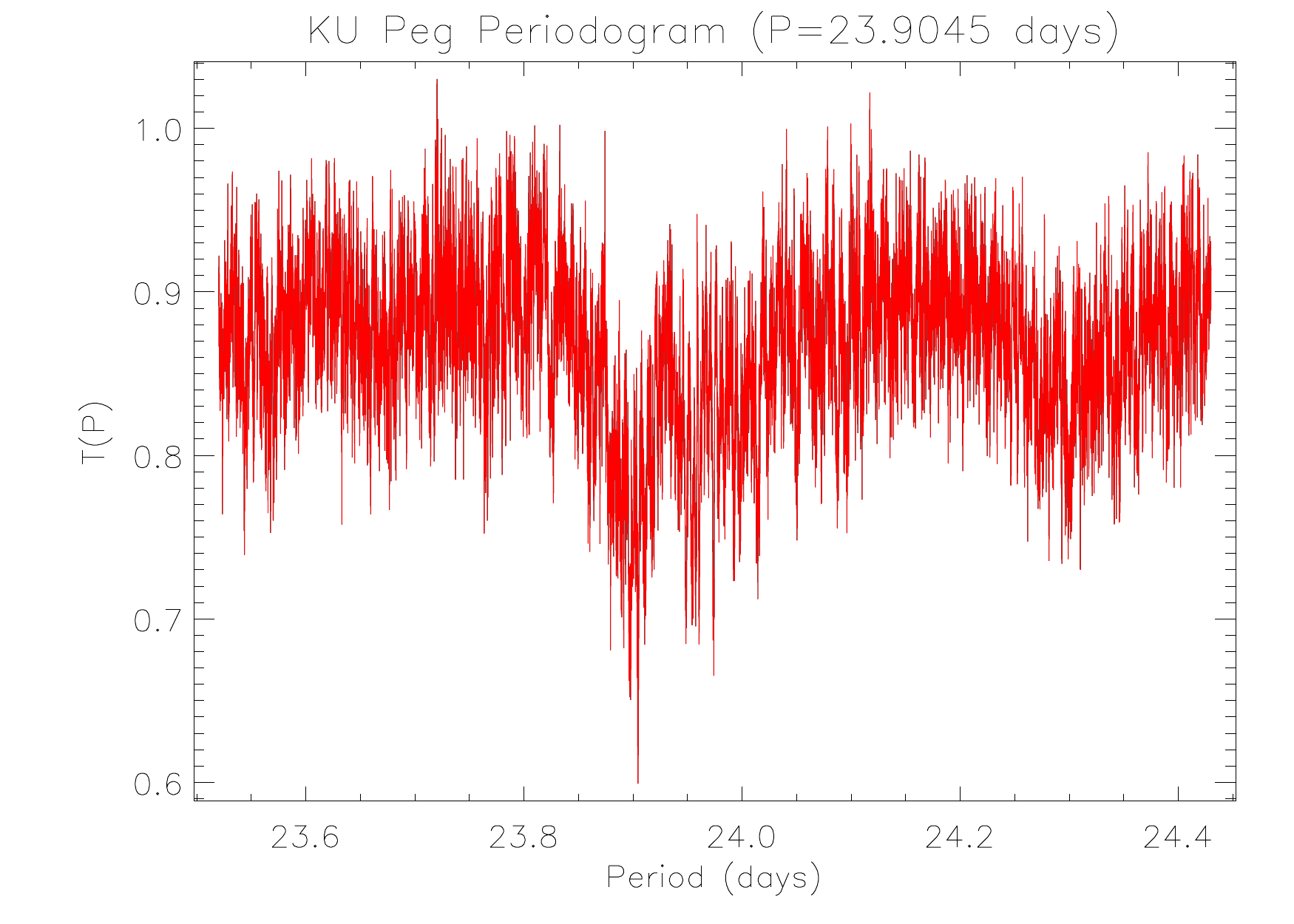}
\caption{String-length Lafler-Kinman periodogram from APT $V$-band data covering $\approx$18~years. Its best-fit
period of 23.9045\,d is interpreted to be the rotation period of the star.}
\label{sllk}
\end{figure}

% ----------------------- F2
\begin{figure}[t]
\includegraphics[angle=0,width=1.0\columnwidth]{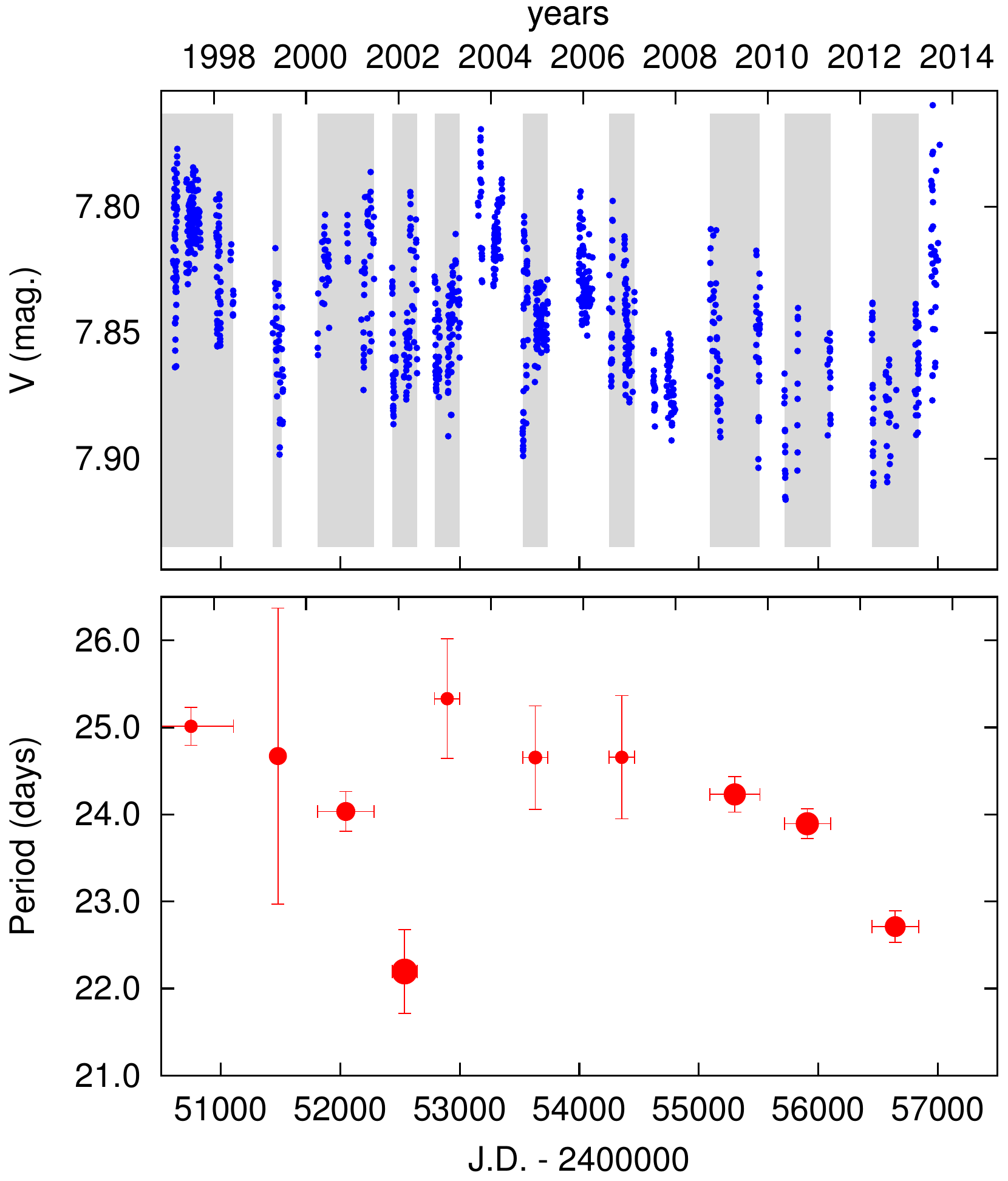}
%\vspace{-5.3cm}
\caption{Johnson $V$ observations (top) and seasonally derived photometric periods (bottom) of \kup.
Each grey band indicates the time range of the data combined for the period determination in the lower panel. In the bottom panel the size of a dot is proportional to the amplitude of the frequency peak in the power spectrum, the vertical error bar represents the uncertainty of the period determination. The horizontal bar just indicates the time coverage.}
\label{seasonal_phot}
\end{figure}

\section{Refinement of the photometric period}\label{prot}

Our extended photometric coverage of more than 18 years allows us to derive a more precise rotation period.
For the period determination we apply the string-length search using the Lafler-Kinman statistic
(hereafter SLLK-method, \citealt{2002A&A...386..763C}). This method phases the light curves with different periods
and selects the period giving the smoothest light curve as the correct one. The SLLK-method is particularly
useful for finding periods in non-sinusoidal data as compared to standard Fourier analysis. Fig.~\ref{sllk} shows
the resulting periodogram and suggests a long-term average photometric period of $P_{\rm phot} = P_{\rm rot} = 23.9045\pm0.0014$\,d
which we adopt as the best representation of the rotation period of \kup . All phase calculations in this
paper use this period with the following ephemeris,
\begin{equation}\label{eq1}
{\rm HJD} = 2\,450\,385.5 + 23.9045\times E,
\end{equation}
where the arbitrarily chosen zero point at HJD\,2\,450\,385.5 is taken from Paper~I for consistency reasons.

In Paper~I a photometric period of $24.96\pm0.04$\,d was derived from the first three years of APT $V$-band data. Its difference of almost one day, i.e. 25$\sigma$, is not only due to measuring errors but due to changing spot locations in combination with differential surface rotation. Photometric periods derived for individual observing seasons on a differentially rotating star are expected to differ from the period from longer-term data (see the review by \citealt{2009A&ARv..17..251S}, and the many references therein). From the range of seasonal periods a rough estimation can be given for the average DR as first introduced by \citet{1972PASP...84..323H}.

In order to determine seasonal periods for \kup, we first select adjacent light curves with the criterion that they
have similar amplitude, mean brightness, and overall shape. Then, a period is determined for each of these (10)
seasonal subsets using our standard Fourier-transformation based frequency analyzer program MuFrAn \citep{2004ESASP.559..396C}.
Fig.~\ref{seasonal_phot} shows the long-term $V$-band APT data along with the subset's time ranges and the period results.
Note that in some cases the light curves were not suitable for deriving a reliable period because of their
large scatter compared to the actual amplitude and/or their insufficient length.
Table~\ref{Tab2} lists the seasonal periods and their errors, which are estimated by increasing the residual scatter of the nonlinear
least-squares solutions some degree, which corresponds to 10\% of the photometric accuracy \citep[cf.][]{2003A&A...410..685O}.

We estimate a DR shear parameter $|\alpha|\gtrsim\Delta P_{\rm phot}/\overline{P}$, where $\Delta P_{\rm phot}$ is the
full range of the seasonal period, while $\overline{P}$ is the long-term average. From the values listed in Table~\ref{Tab2},
we obtain $|\alpha|\gtrsim0.13\pm0.05$. This result is similar to the
value of +0.09 determined in Paper~I from spectroscopy, but considering the errors of either the photometric or the
spectroscopic methods,
it is also in agreement with the corrected value of +0.03 proposed later by \citet{2005AN....326..287W}.
Note, however, that such a method based simply
on photometric data does not allow to determine
the sign of the DR parameter $\alpha$ but only the amount of the shear \citep[but see][]{2015A&A...576A..15R}.

% ---------------------------- T1
\begin{table}
 \centering%%%
\caption{Seasonal periods from our long-term APT photometry.}
\label{Tab2}
%\begin{footnotesize}
\begin{tabular}{c c c}
\hline\noalign{\smallskip}
Seasonal & $P_{\rm phot}$ & $\sigma_{P_{\rm phot}}$ \\
mid-HJD& [d] & [d] \\
\hline\hline
\noalign{\smallskip}
2\,450\,750 &   25.01 &   0.22\\
2\,451\,478 &   24.67 &   1.71\\
2\,452\,046 &   24.04 &   0.23\\
2\,452\,538 &   22.20 &   0.48\\
2\,452\,894 &   25.33 &   0.69\\
2\,453\,631 &   24.66 &   0.59\\
2\,454\,356 &   24.66 &   0.71\\
2\,455\,302 &   24.23 &   0.20\\
2\,455\,908 &   23.89 &   0.17\\
2\,456\,645 &   22.71 &   0.18\\
\hline
 \end{tabular}
\end{table}

\section{Doppler images for 2006--2011}\label{di}

\subsection{Astrophysical parameters of \kup}\label{astrop}

In this section some of the astrophysical parameters are refined with respect to our earlier determinations in Paper~I \citep{2001A&A...373..974W}. Among these are the effective temperature ($T_{\rm eff}$), the surface gravity ($\log g$), the metallicity ([Fe/H]), and the projected rotational velocity ($v\sin i$). We employ our SES spectra and the spectrum-synthesis code PARSES
\citep{2004AN....325..604A,2013POBeo..92..169J}, which is implemented in the standard STELLA-SES data-reduction pipeline \citep{2008SPIE.7019E..0LW}. A grid of synthetic ATLAS-9 spectra tailored to the stellar parameters of \kupt and for up to 40 \'echelle orders around 500--750\,nm
were chosen and the result per spectral order combined on the basis of a weighted least-squares minimization. The average and the standard deviations then constitute our final values and their internal precisions. We found $T_{\rm eff}=4440\pm10$\,K, $\log g$=2.0$\pm$0.1, $v\sin i$=29.4$\pm$1.1\,\kms\  and [Fe/H]=$-0.37\pm0.02$ with a microturbulence of 1.8\kms\ and a prefixed value for the macroturbulence of 3\kms . Note again that the errors are
internal errors. External errors are difficult to obtain for spotted stars with broadened line profiles but from past experience we estimate 70\,K, 0.2\,dex, and 0.1\,dex for $T_{\rm eff}$, $\log g$, and [Fe/H], respectively. We note that $T_{\rm eff}$ is lower by 260\,K compared to the value of 4700\,K in Paper~I. However, when taking $(V-I)_{C,{\rm br}}=1\fm21\pm0\fm12$ for the brightest (=unspotted) magnitude $V_{\rm br}$ of 7\fm760$\pm$0.043, observed only just recently (see Fig.~\ref{seasonal_phot}), a similarly low value of $T_{\rm eff}$ of $4385\pm20$\,K is obtained using the color index vs. temperature calibration by \citealt{2011ApJS..193....1W}. Moreover, taking $B-V=1\fm13$ \citep{2000A&AS..142..275S} together with [Fe/H]=$-0.37$ and using the metallicity-dependent $T_{\rm eff}$-color calibrations by \citet{2015MNRAS.454.2863H} yields  $T_{\rm eff}=4475\pm83$\,K, i.e., again a significantly lower value compared to that in Paper~I, but in alignment with the aforementioned determination from $(V-I)_{C}$. 

Since this new set of fundamental parameters is in contrast with the ones by previous studies
\citep[cf.][]{2009A&A...504.1011L,2015A&A...574A..90A} we carried out a comparative study by using
the spectrum synthesis code SPECTRUM by \mbox{R. Gray} ({\tt www.appstate.edu/$\sim$grayro/spectrum/spectrum.html}). We calculated synthetic spectra from our new parameters ($T_{\rm eff}=4440$\,K, $\log g$=2.0, $v\sin i$=29.4\,\kms, [Fe/H]=$-0.37$, and micro- and macroturbulences of $\xi_{\rm mic}$=1.8\,\kms\ and $\xi_{\rm mac}$=3.0\,\kms, respectively) and from the old parameter set taken from \citet{2009A&A...504.1011L}, i.e., $T_{\rm eff}=5000$\,K, $\log g$=3.0, $v\sin i$=27.1\,\kms, [Fe/H]=$-0.15$, $\xi_{\rm mic}$=2.0\,\kms, $\xi_{\rm mac}$=3.0\,\kms (this latter assumed). The synthetic datasets were compared to the observations (the average of 94 high quality spectra) over the 5950--6510\AA\ spectral range. We found at all times that, in terms of goodness-of-fit values, the synthetic spectra from our new input parameters fitted slightly better the observations. Moreover, the comparisons were extended to some orders between 5000--5600\AA, which resulted in similarly better fits for the new parameters, however, with larger rms values due to the ambiguous continuum setting and the increasing line concentration. Accordingly, we beleive that our new fundamental parameters for \kupt
with lower $T_{\rm eff}$ and lower metallicity are more accurate and more consistent. This is strengthened by the color-temperature calibrations from either $B-V$ or $V-I$ measurements, suggesting effective temperatures lower than 5000\,K by 250--600\,K, depending on the calibration method used \citep[cf.][]{1996ApJ...469..355F,2006A&A...452.1021K,2015MNRAS.454.2863H}. In addition, our lower metallicity agrees better with the photometric metallicities of $-0.50$ and $-0.39$ determined by \citet{1993AJ....106...80E}. Finally we note, that most of the spectral type and luminosity classifications of \kupt in the literature refer to one original determination of G8II by \citet{1956PDDO....2..107H} based on the `general appearence' of objective prism spectra covering much narrower spectral range than ours.

The rotation period from photometry combined with the projected rotational velocity of 29.4$\pm$1.1\,\kms\ from PARSES, and the 50\degr$\pm$10\degr\ inclination angle taken from Paper~I, leads to the most likely stellar radius of $R=18.1^{+4.3}_{-2.8}$~R$_{\odot}$. For $T_{\rm eff}$ of 4440\,K this radius is in good agreement with the expected size of a standard K2III giant star \citep{1996AJ....111.1705D}. The bolometric magnitude from $R^2T_{\rm eff}^4$ is then $M_{\rm bol} = -0\fm39^{+0.37}_{-0.46}$ (adopting $M_{{\rm bol,}\odot}=4\fm74$).

The \emph{Hipparcos} distance of $202^{+27}_{-22}$\,pc \citep{2007A&A...474..653V} combined with $V_{\rm br}$ and an interstellar extinction of $A_V=0\fm42$ (cf. Paper~I) as well as a bolometric correction of $BC=-0\fm64$ from \citet{1996ApJ...469..355F} yields $M_{\rm bol}= 0\fm17^{+0.30}_{-0.32}$. This value is only insignificantly larger than the value calculated from the radius and the effective temperature. As a tradeoff we take $M_{\rm bol}=-0\fm11\pm0\fm28$, i.e., the mean of the two different values, which yields a luminosity of $L=87^{+28}_{-21}L_{\odot}$ for \kup.

\begin{figure}[t]
\includegraphics[angle=0,width=1.0\columnwidth]{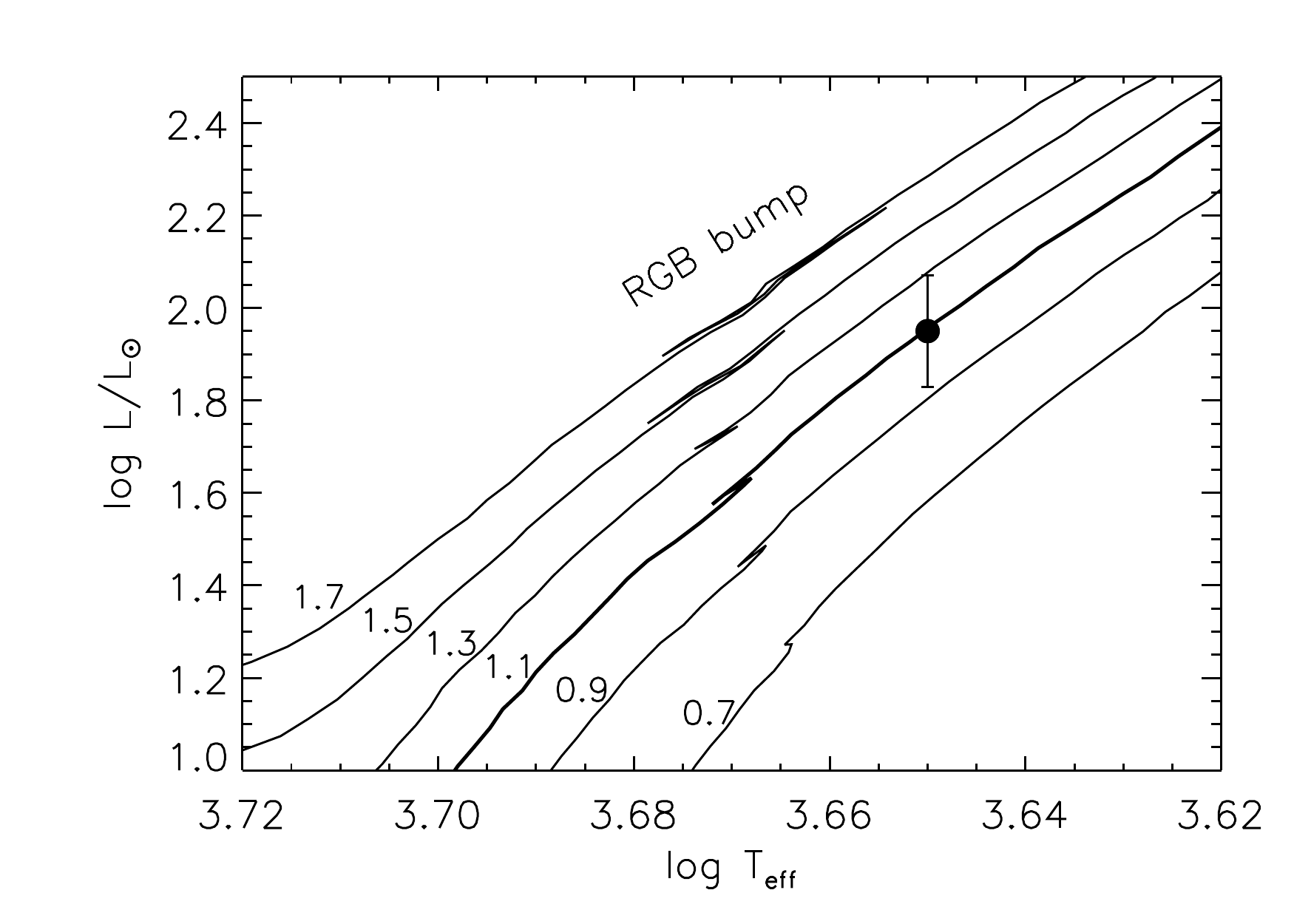}
\caption{Stellar evolutionary tracks around the RGB bump from \cite{2008A&A...484..815B} for $Z=0.008$ together with the position of \kupt (dot with error bar). The numbers indicate the corresponding masses in M$_{\odot}$. It suggests $M=1.1\pm 0.1 M_{\odot}$ as the most likely mass for \kup .}
\label{hrd}
\end{figure}

Fig.~\ref{hrd} shows the position of \kupt in the Hertzsprung-Russell diagram (HRD) together with stellar evolutionary tracks for  $Z$=0.008 of \cite{2008A&A...484..815B}. To determine the mass and age of \kup, a trilinear interpolation within the three-dimensional space ($L$, $T_{\rm eff}$, [Fe/H]) based on a Monte Carlo method is applied \citep{2015A&A...578A.101K}.
The values obtained are a mass of 1.1\,$\pm$\,0.1~$M_{\odot}$ and an age of 7.6\,$\pm$\,2.9~Gyr, assuming the metallicity from PARSES. Note, that the mass is about the half of the former value in Paper~I, thus increasing the age by a factor of $\approx$9. The refined absolute dimensions
and astrophysical quantities summarized in Table~\ref{T3} are more consistent and with generally smaller error bars when compared to Paper~I.

% ----------------------- F3

%-------------------------   T2
\begin{table}[b]
\caption{The astrophysical properties of \kup.} \label{T3}
 \centering
 \begin{tabular}{lll}
  \hline\noalign{\smallskip}
  Parameter               &  Value \\
%  \noalign{\smallskip}
  \hline\hline
  \noalign{\smallskip}
  Spectral type            & K2III  \\
  Distance$_{\rm HIP}$ [pc]    &  $202^{+27}_{-22}$\\
  $V_{\rm br}$    [mag]          & $7\fm760\pm0\fm043$ \\
%   $(B-V)_{\rm HIP}$     [mag]         & 1\fm134 \\
  $(V-I)_{C,{\rm br}}$  [mag]         & $1\fm21\pm0\fm12$ \\
  $M_{\rm bol}$     [mag]        & $-0\fm11\pm0\fm28$ \\
  Luminosity [$\log\frac{L}{L_{\odot}}$]         & $1.94\pm0.12$  \\
  $\log g$ [cgs]        &          $ 2.0\pm0.1$ \\
  $T_{\rm eff}$ [K]           &           $4440\pm10 $    \\
  $v\sin i$ [km\,s$^{-1}$]           &         $29.4\pm1.1$ \\
  Rotation period [d]   &           $23.9045\pm0.0014 $ \\
  Inclination  [\degr]            &       $50\pm10$ \\
  Radius      [$R_{\odot}$]           &      $18.1^{+4.3}_{-2.8}$   \\
  Mass          [$M_{\odot}$]           & $1.1\pm0.1$   \\
  Microturbulence  [km\,s$^{-1}$] & $1.8\pm 0.1$ \\
  Macroturbulence  [km\,s$^{-1}$] & 3.0 \\
  Metallicity [Fe/H] &  $-0.37\pm 0.02$ \\
  NLTE Li abundance (log)   & 0.1$\pm$0.1  \\
\hline
 \end{tabular}
\end{table}

\subsection{The STELLA data subsets}

The spectroscopic data cover up to three consecutive stellar rotations with fairly good phase sampling
in all of the five observing seasons. The detailed time stamps (mid-HJDs) of the data subsets used for the Doppler reconstructions are listed in Table~\ref{Tab4}. For the 2006 and the 2008 seasons three
consecutive data subsets are formed (dubbed S1, S2, and S3, respectively), each covering one single
stellar rotation, i.e., one by one suitable for Doppler reconstruction. For the 2009 and 2011 seasons,
two consecutive subsets are  formed (dubbed S1 and S2, respectively), while for season 2010 only one data set is available.

% ---------------------------- T3
\begin{table*}
 \centering
\caption{Data subsets for the 11 individual Doppler reconstructions.}
\label{Tab4}
%\begin{footnotesize}
\begin{tabular}{c c c c c c c }
\hline\noalign{\smallskip}
Year & Subset & Mid-HJD & Mid-date & Number & Data length & Data length \\
 &  &  & [decimal year] & of spectra & [days] & in $P_{\rm rot}$  \\
\hline
\hline
\noalign{\smallskip}
2006 & S1  & 2\,453\,959.67 & 2006.61  & 19 & 23.24 & 0.97  \\
 & S2  &  2\,453\,980.72   & 2006.67 & 20 & 23.21 & 0.97 \\
 & S3  &  2\,454\,013.05  & 2006.76  & 21 & 22.98 & 0.96 \\
\hline
\noalign{\smallskip}
2008 & S1 & 2\,454\,629.65   & 2008.45 & 22 & 22.92 & 0.96 \\
 & S2  &  2\,454\,658.13  & 2008.52 & 22 & 23.82 & 1.00\\
 & S3  & 2\,454\,684.34  & 2008.60 & 21 & 22.95 & 0.96 \\
\hline
\noalign{\smallskip}
2009 & S1  &  2\,455\,031.29 & 2009.54 & 12 & 22.92 & 0.96  \\
 & S2  & 2\,455\,055.46   & 2009.61 & 19 & 21.80 & 0.91 \\
\hline
\noalign{\smallskip}
2010 & S1  & 2\,455\,448.68  & 2010.69 & 12 & 20.92 & 0.88 \\
\hline
\noalign{\smallskip}
2011 & S1   & 2\,455\,731.32  & 2011.46 & 12 & 18.96 & 0.79\\
 & S2  &    2\,455\,750.37   & 2011.52 & 13 & 18.87 & 0.79 \\
\hline
\end{tabular}
%\end{footnotesize}
\end{table*}

\subsection{The image reconstruction code iMap}

Our DI+ZDI code \emph{iMap} used in this work performs multi-line inversion for a large number of
photospheric line profiles simultaneously \citep{2012A&A...548A..95C}. Note that in this particular case the DI code is used only.
In the case of \kupt 40 suitable
absorption lines, mostly Fe\,{\sc i}, were chosen between 5000--6750 \AA\ \citep{2015A&A...578A.101K}. In the course of
the selection, the line depth, the blends, the continuum level, and the
temperature sensitivity were taken into consideration.

% ---------------------------- F4
\begin{figure*}[thb]
%\begin{center}
\vspace{2.5cm}{\hspace{0.5cm}\huge{S1}}
\vspace{-2.5cm}

\hspace{2.00cm}\includegraphics[angle=0,width=1.7\columnwidth]{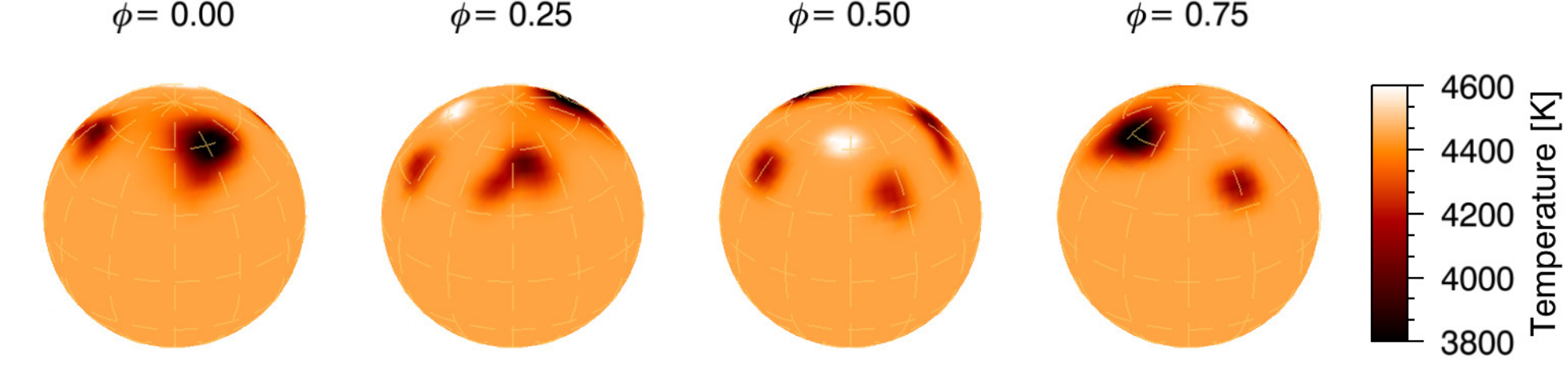}

\vspace{2.5cm}{\hspace{0.5cm}\huge{S2}}
\vspace{-2.5cm}

\hspace{2.00cm}\includegraphics[angle=0,width=1.7\columnwidth]{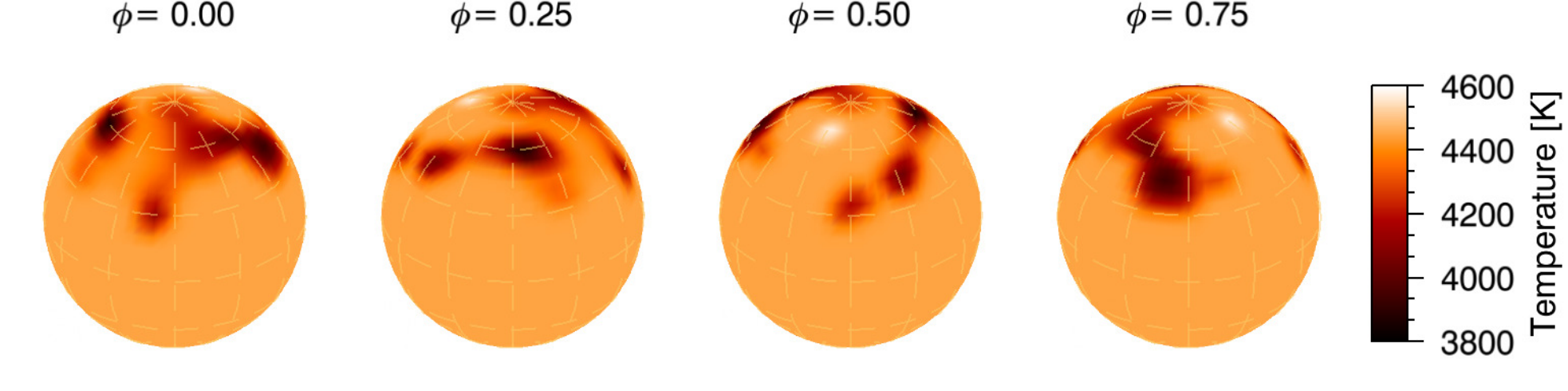}

\vspace{2.5cm}{\hspace{0.5cm}\huge{S3}}
\vspace{-2.5cm}

\hspace{2.00cm}\includegraphics[angle=0,width=1.7\columnwidth]{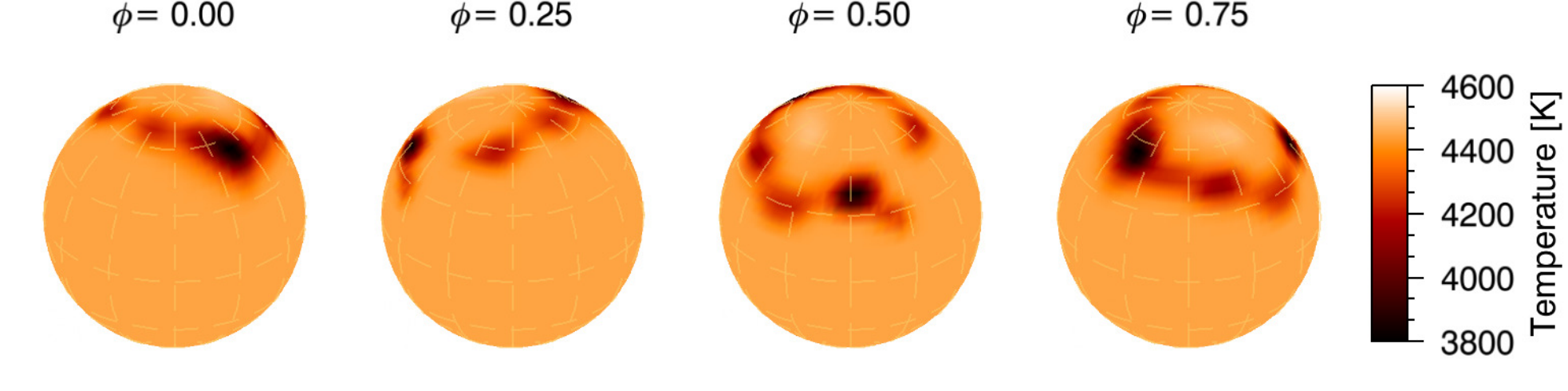}
%\end{center}
\caption{Doppler images of \kupt for the three data sets S1, S2, and S3 in
2006. The corresponding time stamps in mid-dates are 2006.61,  2006.67, and  2006.76, respectively.
The maps are shown in four spherical projections with the corresponding temperature scale. Rotational phase is indicated on the top. }
\label{2006di}
\end{figure*}

For the line profile calculation \emph{iMap} solves the radiative transfer using an artificial
neural network \citep{2008A&A...488..781C}. Individual atomic line parameters are taken from the VALD line database
\citep{1999A&AS..138..119K}. The code uses Kurucz model atmospheres \citep{2004astro.ph..5087C}
which are interpolated for each desired temperature, gravity and metallicity.
The typical ill-posed nature of the surface inversion is tackled with an iterative regularization
based on a Landweber algorithm \citep{2012A&A...548A..95C}. Therefore, no additional
constraints are imposed in the image domain. The surface grid is set to a $5^{\circ} \times 5^{\circ}$
equal-degree partition. For each surface segment the full radiative transfer of all involved line profiles
are calculated under the actual effective temperature and atmospheric model. The line profile discrepancy
is reduced by adjusting the surface temperature of each segment according to the local temperature
gradient information until the minimum $\chi^2$ is reached.

% ---------------------------- F5
\begin{figure*}[pth]
%\begin{center}
\vspace{2.5cm}{\hspace{0.5cm}\huge{S1}}
\vspace{-2.5cm}

\hspace{2.00cm}\includegraphics[angle=0,width=1.7\columnwidth]{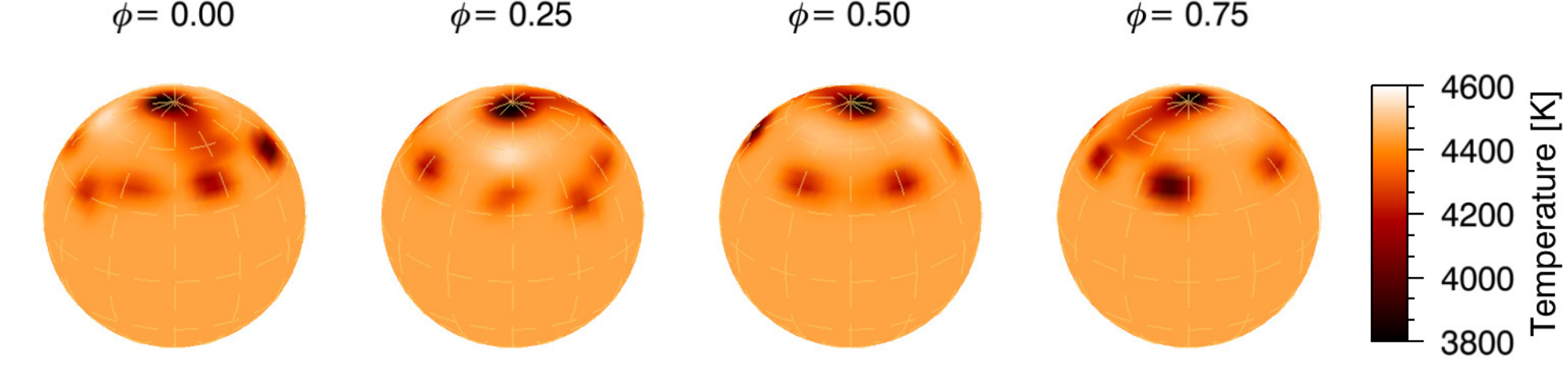}

\vspace{2.5cm}{\hspace{0.5cm}\huge{S2}}
\vspace{-2.5cm}

\hspace{2.00cm}\includegraphics[angle=0,width=1.7\columnwidth]{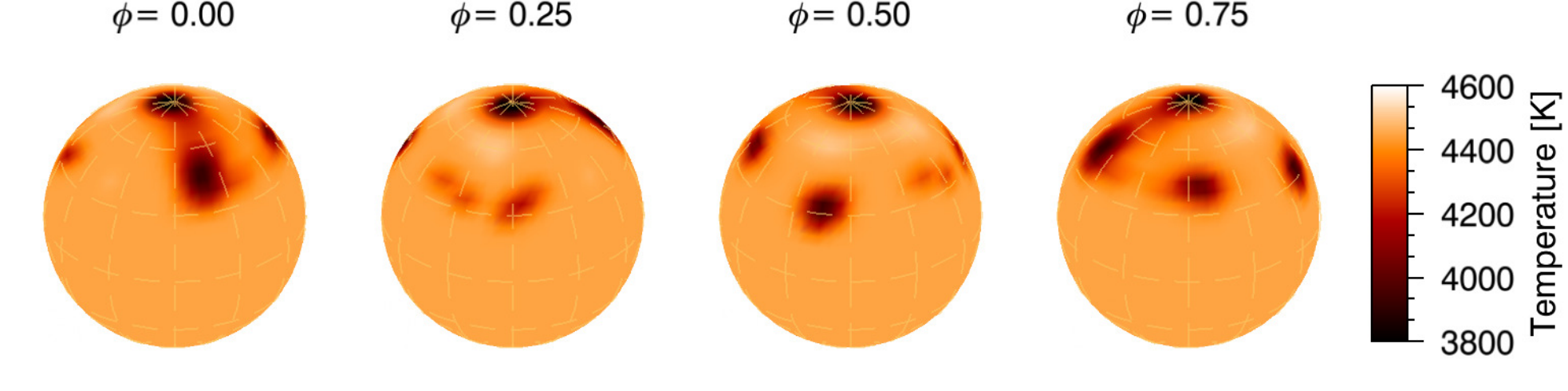}

\vspace{2.5cm}{\hspace{0.5cm}\huge{S3}}
\vspace{-2.5cm}

\hspace{2.00cm}\includegraphics[angle=0,width=1.7\columnwidth]{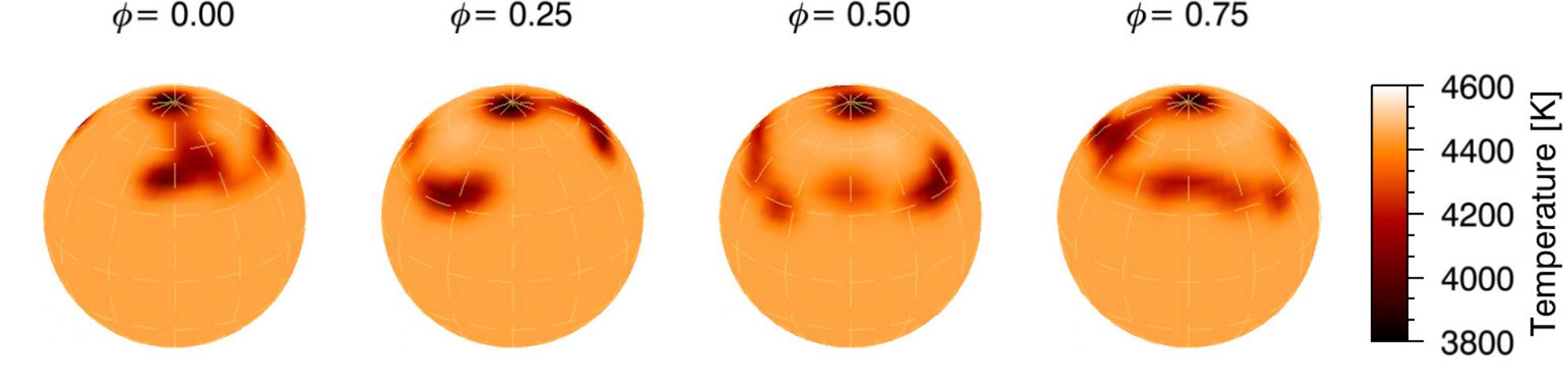}
%\end{center}
\caption{Doppler images of \kupt for the three data sets S1, S2 and S3 in
2008. The corresponding  mid-dates are 2008.45,  2008.52, and  2008.60, respectively.
Otherwise as in Fig.~\ref{2006di}.}
\label{2008di}
\end{figure*}

% ---------------------------- F6
\begin{figure*}[pthb]
%\begin{center}
\vspace{2.5cm}{\hspace{0.5cm}\huge{S1}}
\vspace{-2.5cm}

\hspace{2.00cm}\includegraphics[angle=0,width=1.7\columnwidth]{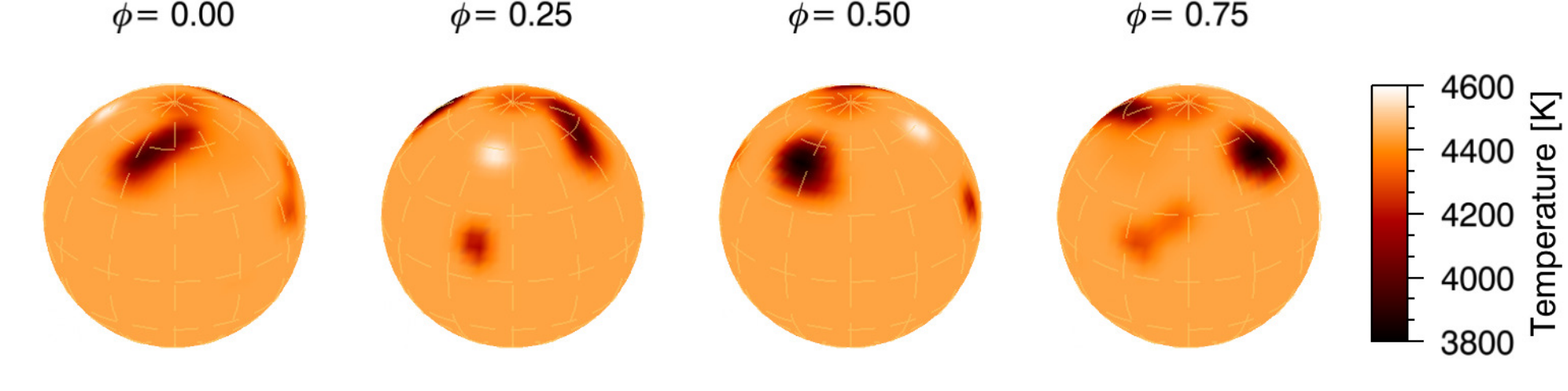}

\vspace{2.5cm}{\hspace{0.5cm}\huge{S2}}
\vspace{-2.5cm}

\hspace{2.00cm}\includegraphics[angle=0,width=1.7\columnwidth]{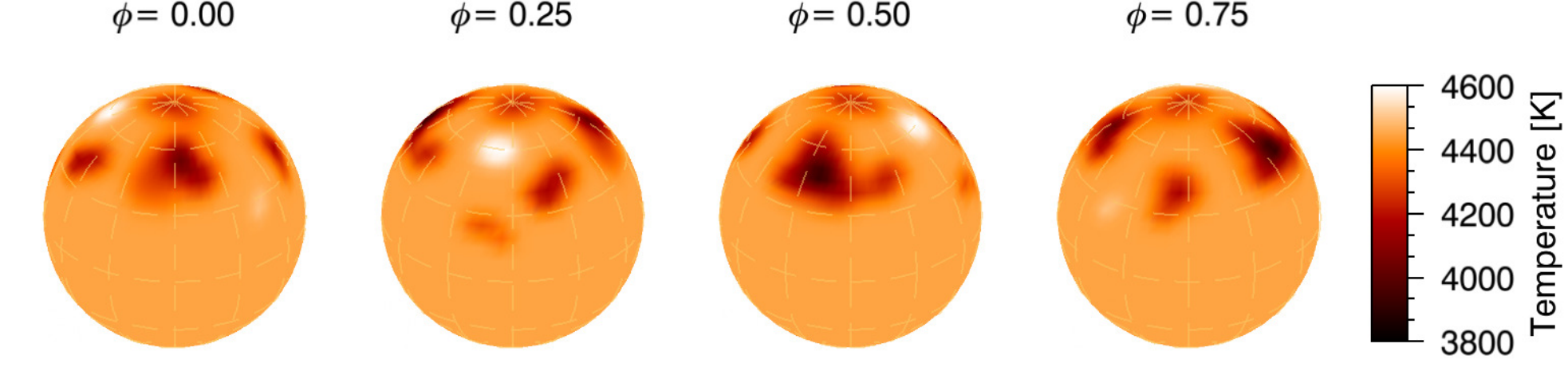}
%\end{center}
\caption{Doppler images of \kupt for the two datasets S1 and S2 in
2009. The corresponding mid-dates are 2009.54, and 2009.61, respectively.
Otherwise as in Fig.~\ref{2006di}.}
\label{2009di}
\end{figure*}

% ---------------------------- F7
\begin{figure*}[thb]
%\begin{center}
\vspace{2.5cm}{\hspace{0.5cm}\huge{S1}}
\vspace{-2.5cm}

\hspace{2.00cm}\includegraphics[angle=0,width=1.7\columnwidth]{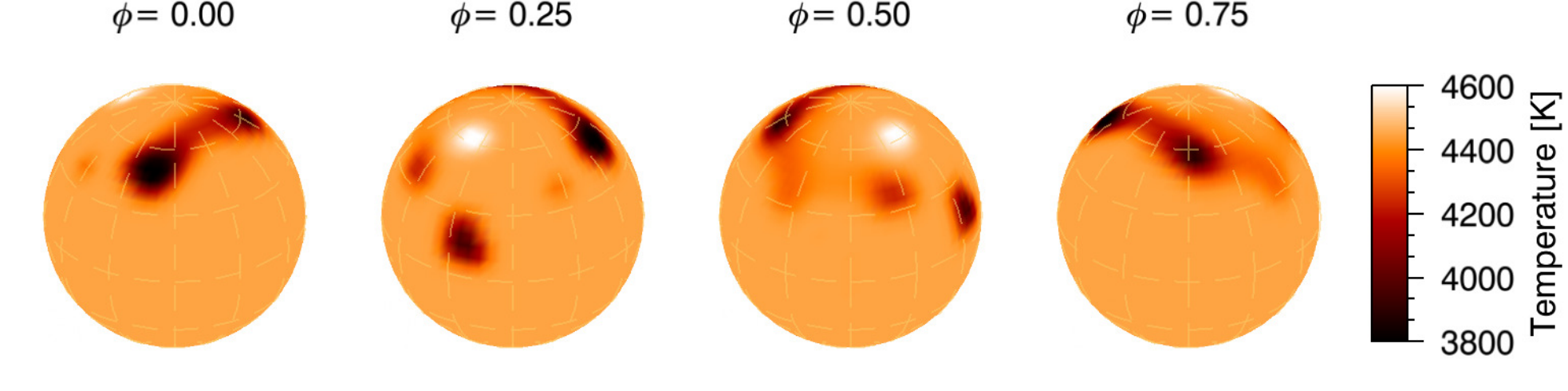}
%\end{center}
\caption{Doppler image of \kupt for the only available dataset (S1/2010) in 2010. The corresponding mid-date is 2010.69. Otherwise as in Fig.~\ref{2006di}.}
\label{2010di}
\end{figure*}

% ---------------------------- F8
\begin{figure*}[thb]
%\begin{center}
\vspace{2.5cm}{\hspace{0.5cm}\huge{S1}}
\vspace{-2.5cm}

\hspace{2.00cm}\includegraphics[angle=0,width=1.7\columnwidth]{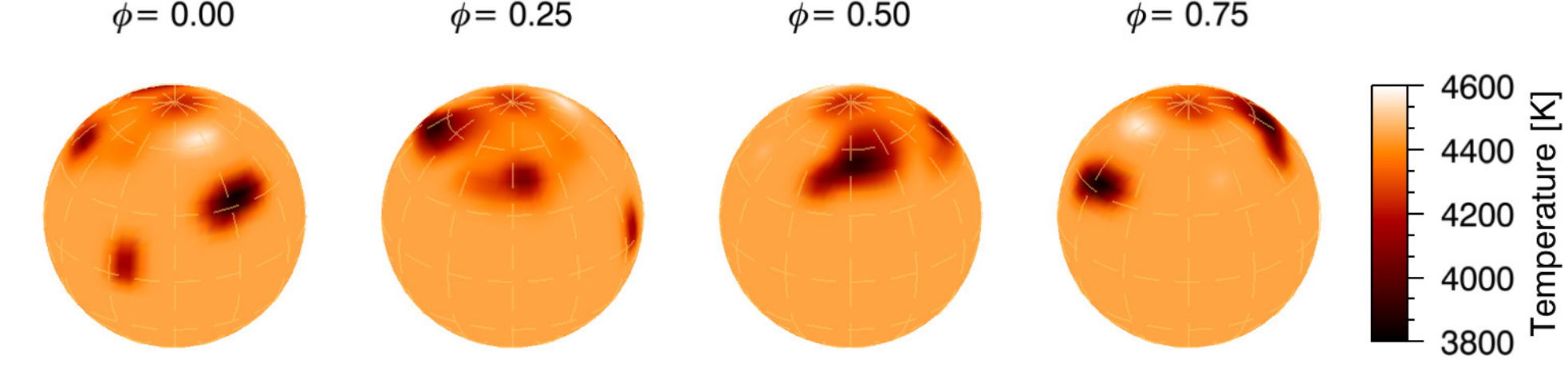}

\vspace{2.5cm}{\hspace{0.5cm}\huge{S2}}
\vspace{-2.5cm}

\hspace{2.00cm}\includegraphics[angle=0,width=1.7\columnwidth]{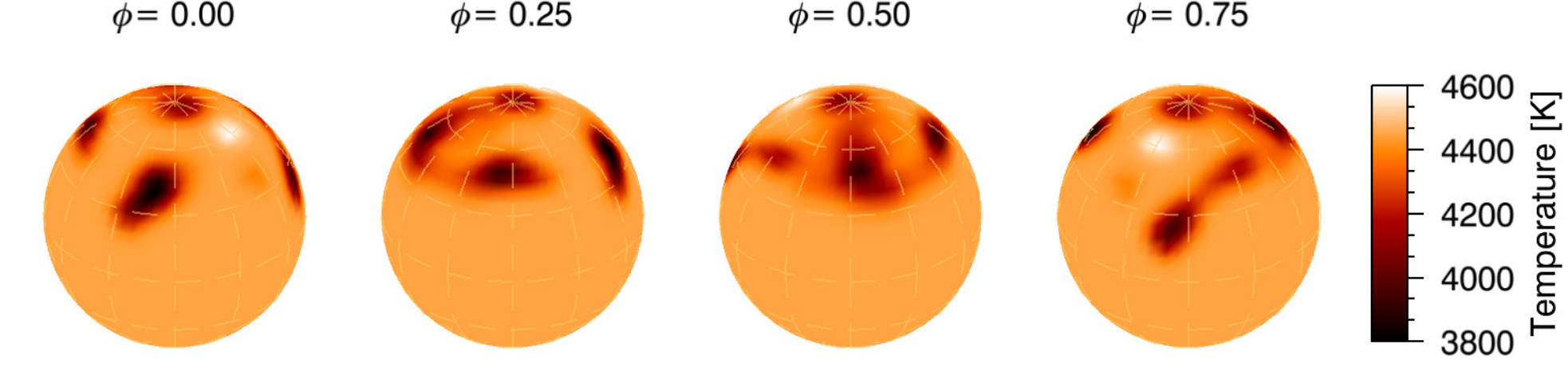}
%\end{center}
\caption{Doppler images of \kupt for the two available datasets (S1 and S2) in
2011. The corresponding mid-dates are 2011.46 and 2011.52, respectively.
Otherwise as in Fig.~\ref{2006di}.}
\label{2011di}
\end{figure*}

\subsection{Doppler image reconstructions}

Our Doppler reconstructions for \kupt in Figs.~\ref{2006di}--\ref{2011di} reveal spots mostly at mid to high latitudes, sometimes covering the visible pole and sometimes a single spot appears at low latitudes. All in all, the images very much resemble the first Doppler images in Paper~I. The temperature of the coolest spots range between $\approx$3650--3850\,K, i.e., cooler by $\approx$800\,K on average than the photosphere.

The three maps in 2006 (S1, S2 and S3 in Fig.~\ref{2006di}) show spots centered at latitudes between 30\degr--80\degr. No polar spot is seen. There is also a bright spot at $\phi\approx0.5$ in S1 with a temperature of $\approx$4700\,K, i.e. $\approx$250\,K warmer than the surrounding photosphere. This feature appears also in the subsequent independent images but with decreasing size and contrast. The three maps in this mini time-series already indicate the fast time evolution of the spotted surface. Short-term rearrangements are seen throughout, which finally result in a ring-like structure around the visible pole (seen in S3).

In the 2008 season the most prominent feature that appears in all of the three Doppler reconstructions (S1, S2 and S3 in Fig.~\ref{2008di}) is the cool polar spot cross talking with another spot centered at around 45\degr\ latitude and 15\degr\ longitude (i.e., best seen at $\phi=0.0$). Again, there is a bright spot best seen at $\phi=0.25$. However, its contrast is weaker than the feature in 2006 and it almost completely vanishes in the last image of the time series. Most interestingly, a ring-like structure had formed by the end of this mini time series, similar to the one in the map S3/2006.

The most dominant feature in the two consecutive Doppler maps in 2009 (S1 and S2 in Fig.~\ref{2009di})
is a cool spot at 50\degr\ latitude with a diameter of $\approx$30\degr. The small polar spot seen a year before is still there but is much weaker. Again there is a small bright spot at 60\degr\ latitude (best seen at $\phi=0.25$) which is getting bigger/warmer in the second image. Other, even smaller features are consistently reconstructed in both images and seen to evolve from S1 to S2.

For the one image in 2010 (S1 in Fig.~\ref{2010di}) the polar spot had fully disappeared while a high latitude feature with an elongated bipolar structure remained and is now dominating the surface. A bright spot is also seen near 60\degr\ latitude but is accompanied by a similar sized cool spot along the same iso-radial line on either side of the central meridian, which makes its reality a little bit suspect. One of its cool counterparts is a rather large and very significant spot though, which is a counterargument because it is actually well constrained.

The two consecutive Doppler images in 2011 (S1 and S2 in Fig.~\ref{2011di}) reveal dramatic changes in the spot morphology over just about one stellar rotation. The largest changes are seen at lower latitudes. The one cool spot located at around 45\degr\ latitude and 15\degr\ longitude in S1 seemed to have been shifted by more than 20\degr\ towards increased longitude in S2 or is a product of a merger. The other smaller spot centered at 10\degr\ latitude and 345\degr\ longitude seems to have either disappeared or merged with the spot at 15\degr\ longitude. Meanwhile, a minor displacement of the high latitude bright feature between S1 and S2 is seen in the direction of forward rotation. The most dominant cool spot or spot group in S1 at 180\degr\ longitude (best seen at $\phi=0.5$) seems to have started dissolving or at least stretching towards other nearby spots. A weak polar spot seems to be getting stronger in the second map.

The line profile fits for the altogether 11 Doppler reconstructions in the five observing seasons are plotted in Figs.~\ref{proffits1}--\ref{proffits2} in the Appendix.

\section{Surface differential rotation from Doppler images}\label{ccf}

Time-series Doppler images allow a determination of the surface DR by cross-correlating the consecutive maps with each other. We apply our cross-correlation technique \texttt{ACCORD} \citep{2012A&A...539A..50K}, which combines all the available surface information in order to achieve an intensified signature of DR. In our case, we have a total of 11 Doppler images. In 2006 and 2008 we have 3 consecutive maps, while in 2009 and 2011 we have 2 consecutive maps. Therefore, we are able to create altogether 8 pairs of maps, i.e., S1-S2, S2-S3 and S1-S3 for 2006 and 2008 and S1-S2 for 2009 and 2011. We cross-correlate the corresponding latitude stripes of the paired Doppler images for each
latitude bin of 5\degr-width, obtaining 8 cross-correlation function maps. These correlation maps are then combined in order to recover an average correlation pattern from which we determine the surface DR. For a more detailed description of the \texttt{ACCORD} technique we refer to our recent application in \citet{2015A&A...573A..98K} and the references therein.

\begin{figure}[thb]
\includegraphics[width=1.0\columnwidth]{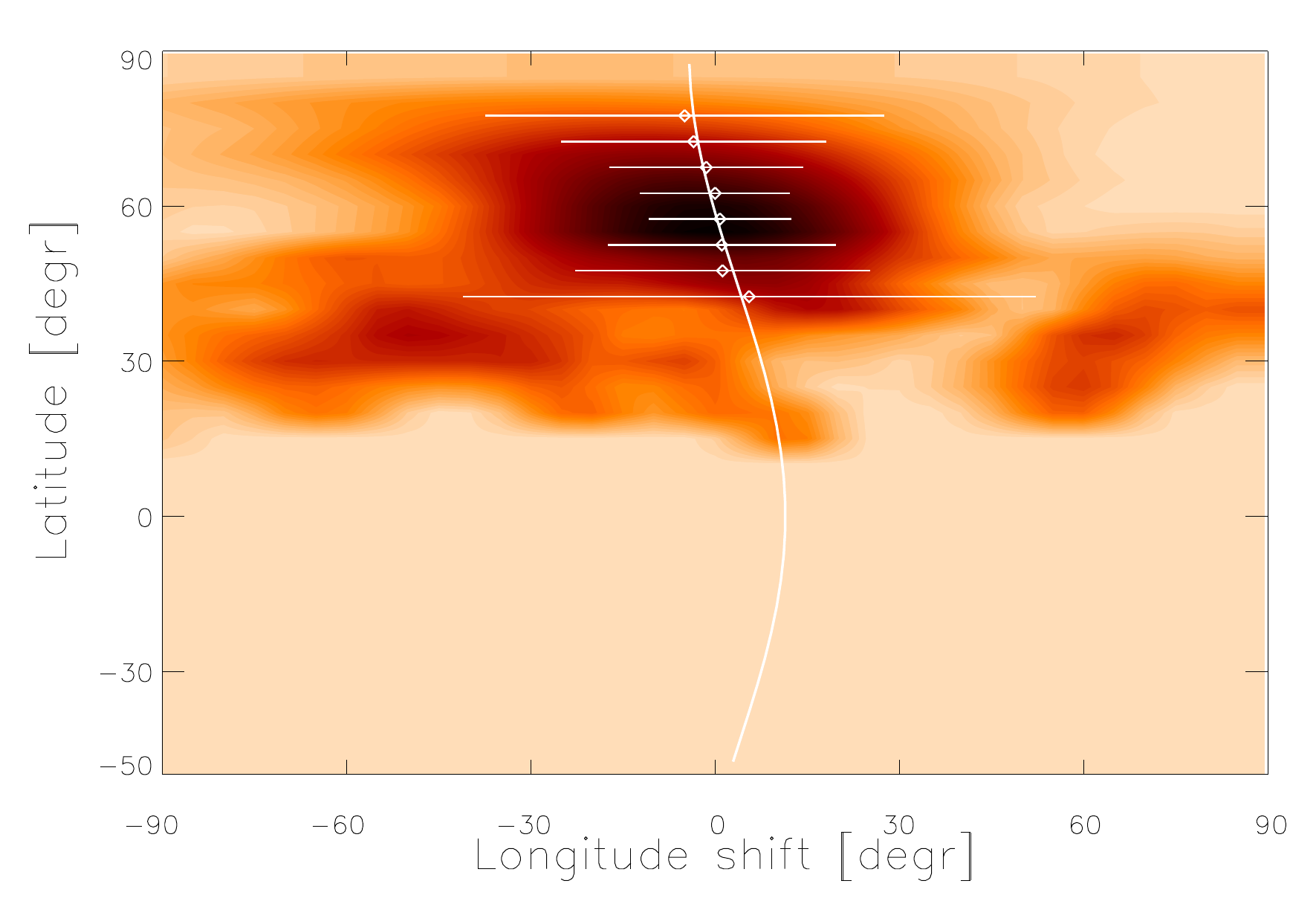}
\caption{Average cross-correlation function map showing the evidence for surface differential rotation. Darker shade represents better correlation. The average longitudinal cross-correlation functions in 5\degr\ bins are fitted by Gaussian curves. Gaussian peaks are indicated by dots, the corresponding Gaussian widths by horizontal lines. The continuous line is the best fit, suggesting solar-type differential rotation with $P_{\rm eq}=23.2051$\,d equatorial period and $\alpha=+0.040$ surface shear.}
\label{ccf_iMap}
\end{figure}

Fig.~\ref{ccf_iMap} shows the average correlation pattern with the best-fit quadratic rotation law. Note that due to the limited appearance of spots at low latitudes, the weak correlation pattern at latitudes below $\approx$\,40\degr\ must be rejected. Yet the dashed-line function clearly represents a solar-type DR. The rotation law in the usual quadratic form takes the shape $\Omega(\beta)=\Omega_{\rm eq}(1-\alpha\sin^2\beta)$, where $\Omega(\beta)$ is the angular velocity
at $\beta$ latitude, $\Omega_{\rm eq}$ is the equatorial angular velocity, while $\alpha=\Delta\Omega/\Omega_{\rm eq}$ is the surface shear coefficient and $\Delta\Omega=\Omega_{\rm eq}-\Omega_{\rm pol}$ is the angular velocity difference between the equator and the pole. The best fit yields $\Omega_{\rm eq}=15.5138$\degr/d (or equivalently $P_{\rm eq}=23.2051$\,d) and $\alpha=+0.040\pm0.006$. This can be converted to a lap time of $\approx$580\,d, i.e., the time needed by the equator to lap the polar regions.
This result is in the order of the empirical estimation of $|\alpha|\approx P_{\rm rot}/360\,{\rm d}$ deduced by \citet{2014SSRv..186..457K} from Doppler imaging studies of either single stars or members of binary systems. Note however,
that the statistically small sample of comparably fast rotating single giants with known surface DR does not allow to
give such a relationship for single giants only.

% ----------------------- F9

\section{Li abundance determination}\label{lithium}

% ----------------------- F10
\begin{figure}[tb]
\includegraphics[width=1.0\columnwidth]{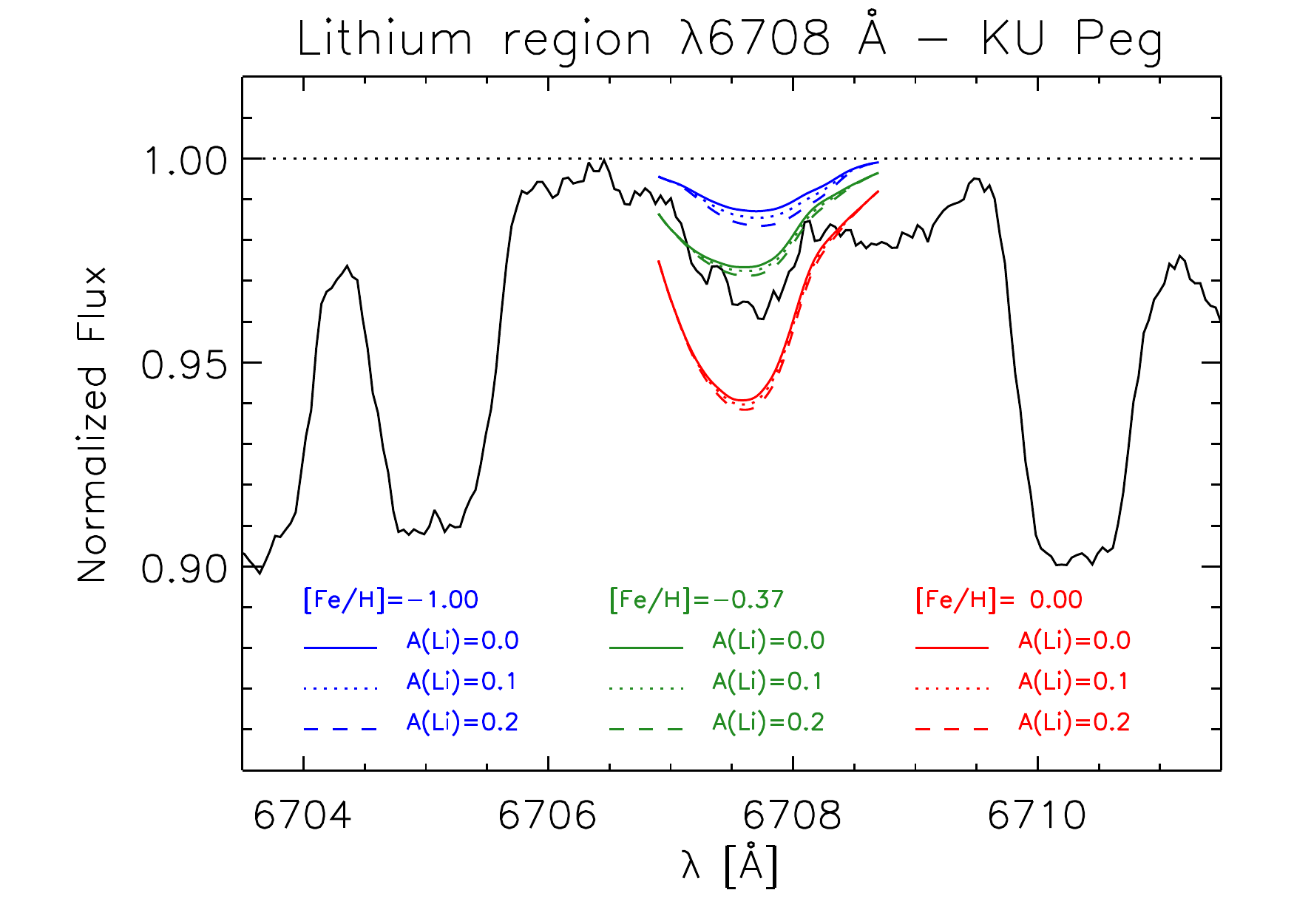}
\caption{Li\,{\sc i} 6708\,\AA\ spectral region of \kup. The observed spectrum is an average of 40 individual exposures taken in July/Aug 2009. The three sets of thin lines are synthetic spectra for three metallicities (see insert, for --1.00, --0.37 and 0.0) and for three Li abundances of $\log n$=0.0, 0.1, and 0.2. Our best estimate with a metallicity of --0.37 suggests an upper Li limit of $\log n$=0.1$\pm$0.1. }
\label{li_spectrum}
\end{figure}

In Fig.~\ref{li_spectrum} the extracted spectral region around the Li\,{\sc i}-6708\,\AA\ line is plotted. Visual inspection of individual exposures does not show any striking evidence of a Li\,{\sc i}-6708\,\AA\ line, hampered by the large $v\sin i$ (30\,\kms ) and the comparably low S/N (on average 100:1 for a single exposure). Only after co-adding 40 spectra from within two months in 2009 (July/Aug) a consistent asymmetry of the Fe\,{\sc i} and CN blends appears. However, the averaging smears the line profile shapes due to the rotationally modulated spot contribution and makes the average line strength appear weaker than it is. This may be a few-\%\ effect in the equivalent width but we believe it can be neglected because the uncertainty of the continuum setting is so much larger. A straight-forward double Gaussian fit to the average spectrum centered at the average Li\,{\sc i}-6708\,\AA\ line wavelength and the Fe\,{\sc i}-6707.43 blend results in a mere $\approx$5\,m\AA\ Li equivalent width. It converts to a logarithmic abundance of $\approx$0.1$\pm$0.1 relative to hydrogen ($\log n$(H)=12.00) with the NLTE tables of \citet{1996A&A...311..961P}; if it is all due to Li and for the case that the models are error free. Such a low Li abundance is not seen on other rapidly-rotating K giants of comparable luminosity and \kupt may be an interesting target also for constraining the Li dredge-up problem.

We also compare our average spectrum with a small grid of forward synthetic Li spectra from 3D model atmospheres. A formal fit of the spectrum was not possible with the current set-up due to the lack of the appropriate 3D  models and the comparable large $v\sin i$ and low S/N of the data. However, Fig.~\ref{li_spectrum} compares three sets of synthetic spectra for metallicities of between solar and ten times less than solar and with logarithmic Li abundances of $A$(Li)=0.0, 0.1, and 0.2, i.e., approximately ten times below the solar value ($\log n_\odot$=1.05$\pm$0.1; \citealt{2009ARA&A..47..481A}). We note that changing the continuum of the observed spectrum by just 1\% has already a 0.2-dex impact on the Li abundance and similar is the case for the metallicity. Nevertheless, the 3D synthetic spectra agree with the equivalent width above and constrain the Li abundance to an upper limit of $\log n$=0.1$\pm$0.2. Its uncertainty is estimated from the case when the metallicity and the continuum location are assumed to be free of error. We adopted the new line list given in \citet{2015AN....336..968C} with a total of 40 spectral lines plus the hyperfine structure of the Li resonance. Line computations were done from 3D CO5BOLD model atmospheres matching the temperature and gravity of \kupt in Table~\ref{T3}.

\section{Summary and discussions}\label{disc}

Our new Doppler image reconstructions of \kupt indicate that its surface spot distribution is indeed very active and dynamic, even when compared to other overactive stars, and so must be the underlying dynamo. According to the theoretical estimation by \citet{2015A&A...574A..90A} the maximum convective turnover time of \kupt should be around $\tau\approx100$\,d. This would yield a moderate Rossby number of $Ro\approx0.24$, indicating that the star operates an $\alpha\Omega$ type dynamo. However, from the time evolution of the spotted surface, considering especially the polar spottedness between 2006-2011, we could only estimate a rough cycle length of a few years. On the other hand, from long-term photometry, such a cycle (of about 2--4 years) can be inferred at a very weak significance level, i.e., not conclusively.

From our cross-correlation study, we derived a solar-like surface DR with a shear of $\alpha=+0.040\pm0.006$ and a lap time of $\approx$580~days. A similar solar-like DR was found in Paper~I with $\alpha=+0.09$ and a corresponding lap time of $\approx$260 days. Note that the higher value in Paper~I came from a cross-correlation of only two consecutive Doppler maps and the use of a less-robust correlation routine which both resulted in a less pronounced correlation pattern. A redetermination of $\alpha$ with a different cross-correlation program but the same data as in Paper~I
by \citet{2005AN....326..287W}, revealed an $\alpha$ of +0.03, in agreement with our new value. In the present paper, we applied a more robust cross-correlation technique for altogether 8 cross-correlation maps and conclude that our new result is much more reliable and has now a reasonable error bar.

The time-series Doppler reconstructions revealed evidence for systematic spot displacements that may be interpreted as evidence for local meridional flows. Examples are the poleward drift of the dominant
feature in Fig.~\ref{2006di} at $\phi=0.25$ or the displacement of the low latitude spot in
Fig.~\ref{2009di} at $\phi=0.25$ (as well as in Fig.~\ref{2011di} at $\phi=0.00$ and at $\phi=0.25$). Compared to the longitudinal displacements due to DR, the latitudinal displacements
are more diffused and much weaker on average. At the same time the overall evolution of the spot distribution is generally more complex than thought and can have spots come and go from one rotation to the next (e.g. in 2011, Fig.~\ref{2011di}).

Rapid rotation of an old, effectively single, evolved star like \kupt remains a challenge for theory. If tidal effects did not play a role in the past of the star, the most likely explanation for its rapid rotation would be that a deepening convective envelope eventually reaches the high angular
momentum material around a fast rotating core, and thus transports high angular momentum material up to the surface on a comparably short convective time scale \citep{1979ApJ...232..531E}. This dredge up must take place before the star evolves up to the bump of the red giant branch. On the other hand, with the expanding envelope an increasing mass loss rate would be expected, which would again mean angular momentum loss. However, excessive mass loss can be excluded by the lack of any IR excess from a comparison of the measured (2MASS) $J$, $H$ and $K$ magnitudes and the color calibrations provided by \citet{2005ApJ...626..465R}.
But out-flowing material can also be coupled to closed surface magnetic fields, generated by a dynamo, this way preventing the star from fast angular momentum loss \citep[cf.][]{2010ApJ...719..299C}.

In any case, a rapidly rotating core on the main sequence is required to explain the spin-up by
angular momentum transport from the deep. However, it is not likely that a star of $1.1\,M_{\odot}$ could provide such a fast rotating core. Moreover, according to \citet{2016A&A...591A..45P} the dredge up may not produce enough acceleration of the surface to be a reasonable explanation at all.

As an alternative scenario, engulfment of one planet (or even more) may explain the rapid rotation \citep{1999MNRAS.308.1133S,2012ApJ...757..109C,2016arXiv160608027P}.
Taken the expression from \citet{2008AJ....135..209M} we estimate the mass of the planet which would spin up the star to be $\approx$1.25\,$M_{\rm J}$. But this would also raise the Li abundance at the surface rather than lowering it, which is found for \kup. Note however, that according to \citet{2016arXiv160303038C} any close giant planet is likely to be engulfed well before the host star would evolve up the RGB. This would explain the low surface Li abundance, since by the end of the dredge up phase the extra Li coming from the planet will be destroyed together with the primeval Li of the stellar envelope.
But anyhow, lithium can be affected by other less known processes too, thus the surface Li measurement itself can hardly account for or disprove any planetary interaction as pointed out by \citet{2016arXiv160608027P}.

The position in the H-R diagram indicates that \kupt is past the RGB luminosity bump. Only low-mass stars that have a highly degenerate He core on the RGB, and later undergo the He flash, evolve through this phase \citep[see][]{2000A&A...359..563C}. At this time extra Li is produced and very high Li abundances are reached \citep[e.g., HD\,233517;][]{2015A&A...574A..31S}. However, this phase is extremely short lived because once the mixing extends deep enough the freshly synthesized Li is quickly destroyed. Immediately before the bump phase (and after the end of the first dredge-up), we expect relatively low Li abundances. The time-scale of the bump for $M=1.1 \mathrm{M}_\sun$ and  $Z$=0.008 (almost equivalent with [Fe/H]=$-$0.37) is $\approx$10\,Myr, the time between the bump and the current position of \kupt is $\approx$40\,Myr according to the models of \citet{2008A&A...484..815B}. This must have been enough time to dilute KU\,Peg's surface Li to basically zero.

\begin{acknowledgements}
We are thankful for the critical remarks by the referee, who helped to improve the manuscript.
Authors from Konkoly Observatory are grateful to the \emph{Hungarian Scientific Research Fund}
for support through grants OTKA K-109276 and OTKA K-113117. This work is supported by the ``Lend\"ulet-2011" Young Researchers' Program of the Hungarian Academy of Sciences.
The authors acknowledge the support of the German \emph{Deut\-sche For\-schungs\-ge\-mein\-schaft, DFG\/} through projects KO~2320/1 and STR645/1. We thank Alessandro Mott for computing the synthetic Li spectra for us.
This publication makes use of data products from the Two Micron All Sky Survey, which is a joint project of the
University of Massachusetts and the Infrared Processing and Analysis Center/California Institute of Technology,
funded by the National Aeronautics and Space Administration and the National Science Foundation.
\end{acknowledgements}

%%%%%%%%%%%%%%%%%
%%%%%%%% references%%%
%%%%%%%%%%%%%%%%%

\bibliography{kovarietal_kupeg}
\bibliographystyle{aa}

% -----------------------------------------------------------------------
%
%   A P P E N D I X  -  online material
%
% -----------------------------------------------------------------------
\begin{appendix}
\section{Observing log and line profile fits}

% ---------------------------- TA1

\begin{table*}
 \centering
\caption{Observing log of STELLA-I SES spectra taken between 2006--2011.
}
\label{Tab1}
\vspace{-2mm}
\begin{footnotesize}
\begin{tabular}{|c c c c | c c c c | c c c c|}
\hline\noalign{\smallskip}
HJD\tablefootmark{a}  & Phase\tablefootmark{b} &  S/N  & Subset/year &HJD\tablefootmark{a}   & Phase\tablefootmark{b} &  S/N  & Subset/year &HJD\tablefootmark{a}   & Phase\tablefootmark{b} &  S/N  & Subset/year\\
\hline
\hline\noalign{\smallskip}
3947.4901   &   0.009   &   40    &   S1/2006 & 4623.6530   &   0.295   &   117    &   S1/2008 &5028.6074   &   0.236   &   122    &   S1/2009\\
3949.4836   &   0.093   &   40    &   S1/2006 & 4624.6476   &   0.337   &   127    &   S1/2008  & 5032.5866   &   0.402   &   120    &   S1/2009 \\
3950.5241   &   0.136   &   35    &   S1/2006 & 4625.6432   &   0.378   &   133    &   S1/2008 &5035.6056   &   0.528   &   80    &   S1/2009\\
3951.5597   &   0.179   &   46    &   S1/2006 & 4626.6788   &   0.422   &   139    &   S1/2008 & 5038.5754   &   0.653   &   77    &   S1/2009\\
3953.4742   &   0.260   &   38    &   S1/2006 & 4627.6493   &   0.462   &   114    &   S1/2008 & 5039.5280   &   0.693   &   120    &   S1/2009\\
3954.4723   &   0.301   &   25    &   S1/2006 &  4628.6382   &   0.504   &   136    &   S1/2008  &   5040.5974   &   0.737   &   139    &   S1/2009\\
3954.5571   &   0.305   &   22    &   S1/2006 &  4629.6326   &   0.545   &   103    &   S1/2008  & 5043.6119   &   0.863   &   105    &   S1/2009\\
3955.5405   &   0.346   &   27    &   S1/2006 &  4630.6139   &   0.586   &   98    &   S1/2008  &  5044.6923   &   0.909   &   76    &   S2/2009\\
3958.5047   &   0.470   &   52    &   S1/2006 &  4631.6384   &   0.629   &   120    &   S1/2008  &   5045.6145   &   0.947   &   85    &   S2/2009\\
3959.4636   &   0.510   &   56    &   S1/2006 &  4632.6346   &   0.671   &   137    &   S1/2008  &  5046.5245   &   0.985   &   122    &   S2/2009\\
3960.5446   &   0.555   &   24    &   S1/2006 &  4634.6920   &   0.757   &   107    &   S1/2008  &  5047.6092   &   0.031   &   144    &   S2/2009\\
3961.7125   &   0.604   &   41    &   S1/2006 &  4635.6335   &   0.796   &   120    &   S1/2008  &  5048.6052   &   0.072   &   145    &   S2/2009\\
3963.4532   &   0.677   &   36    &   S1/2006 &  4636.6556   &   0.839   &   115    &   S1/2008  & 5049.5334   &   0.111   &   126    &   S2/2009\\
3966.4524   &   0.802   &   26    &   S1/2006 &  4638.6139   &   0.921   &   82    &   S1/2008  &  5050.5126   &   0.152   &   116    &   S2/2009\\
3967.4415   &   0.844   &   35    &   S1/2006 &  4639.6265   &   0.963   &   117    &   S1/2008  &   5051.5125   &   0.194   &   126    &   S2/2009\\
3968.4339   &   0.885   &   37    &   S1/2006 &  4640.6610   &   0.007   &   112    &   S1/2008  &   5055.5099   &   0.361   &   135    &   S2/2009\\
3969.4851   &   0.929   &   57    &   S1/2006 &  4641.5962   &   0.046   &   101    &   S1/2008  & 5056.5084   &   0.403   &   115    &   S2/2009\\
3970.4301   &   0.969   &   48    &   S1/2006 &  4645.6956   &   0.217   &   100    &   S2/2008  & 5057.5084   &   0.445   &   141    &   S2/2009\\
3970.7259   &   0.981   &   58    &   S1/2006 &  4646.6213   &   0.256   &   94    &   S2/2008  &    5058.5318   &   0.488   &   135    &   S2/2009\\
3971.4292   &   0.011   &   34    &   S2/2006 &  4649.5961   &   0.380   &   100    &   S2/2008  &  5059.5327   &   0.529   &   129    &   S2/2009\\
3971.7251   &   0.023   &   37    &   S2/2006 &  4650.5906   &   0.422   &   114    &   S2/2008  & 5060.7000   &   0.578   &   80    &   S2/2009\\
3972.4294   &   0.052   &   40    &   S2/2006 &  4651.5976   &   0.464   &   128    &   S2/2008  &   5061.6971   &   0.620   &   134    &   S2/2009\\
3972.7261   &   0.065   &   42    &   S2/2006 &  4652.5943   &   0.506   &   112    &   S2/2008  &  5062.4696   &   0.652   &   125    &   S2/2009\\
3973.4273   &   0.094   &   43    &   S2/2006 &  4653.6199   &   0.549   &   116    &   S2/2008  &  5064.7071   &   0.746   &   85    &   S2/2009\\
3974.7180   &   0.148   &   42    &   S2/2006 &  4654.5914   &   0.589   &   96    &   S2/2008  &  5065.5002   &   0.779   &   71    &   S2/2009\\
3975.4258   &   0.178   &   52    &   S2/2006 &  4655.5802   &   0.631   &   91    &   S2/2008  & 5066.4959   &   0.821   &   130    &   S2/2009\\
3976.6662   &   0.230   &   45    &   S2/2006 &  4656.5973   &   0.673   &   91    &   S2/2008  &   5439.5302   &   0.426   &   86    &   S1/2010\\
3977.6598   &   0.271   &   45    &   S2/2006 &  4657.5933   &   0.715   &   100    &   S2/2008  &  5440.5185   &   0.467   &   123    &   S1/2010\\
3978.4228   &   0.303   &   44    &   S2/2006 &  4658.5764   &   0.756   &   103    &   S2/2008  &  5441.4937   &   0.508   &   123    &   S1/2010\\
3979.7216   &   0.358   &   44    &   S2/2006 &  4659.5886   &   0.798   &   130    &   S2/2008  & 5443.4856   &   0.591   &   106    &   S1/2010\\
3980.4605   &   0.388   &   50    &   S2/2006 &  4660.5857   &   0.840   &   125    &   S2/2008  & 5444.5140   &   0.634   &   112    &   S1/2010\\
3984.6764   &   0.565   &   48    &   S2/2006 &  4661.5311   &   0.880   &   121    &   S2/2008  &   5446.4904   &   0.717   &   72    &   S1/2010\\
3985.6648   &   0.606   &   51    &   S2/2006 &  4662.5817   &   0.924   &   103    &   S2/2008  &  5447.5329   &   0.761   &   91    &   S1/2010\\
3987.4155   &   0.679   &   59    &   S2/2006 &  4663.5849   &   0.966   &   115    &   S2/2008  &   5449.6861   &   0.851   &   110    &   S1/2010\\
3987.6347   &   0.689   &   53    &   S2/2006 &  4665.5716   &   0.049   &   122    &   S2/2008  &  5453.4555   &   0.008   &   64    &   S1/2010\\
3987.6892   &   0.691   &   54    &   S2/2006 &  4666.5779   &   0.091   &   118    &   S2/2008  &  5457.4696   &   0.176   &   116    &   S1/2010\\
3988.4145   &   0.721   &   55    &   S2/2006 &  4667.5328   &   0.131   &   81    &   S2/2008  & 5459.4770   &   0.260   &   69    &   S1/2010\\
3993.4111   &   0.930   &   65    &   S2/2006 &  4668.5316   &   0.173   &   104    &   S2/2008  &   5460.4521   &   0.301   &   112    &   S1/2010\\
3994.6385   &   0.982   &   63    &   S2/2006 &  4669.5129   &   0.214   &   105    &   S2/2008  &  5721.6610   &   0.228   &   124    &   S1/2011\\
3999.4056   &   0.181   &   65    &   S3/2006 &  4673.5035   &   0.381   &   99    &   S3/2008  &  5722.6971   &   0.272   &   146    &   S1/2011\\
4003.6096   &   0.357   &   86    &   S3/2006 &  4674.5176   &   0.423   &   106    &   S3/2008  &  5724.6215   &   0.352   &   116    &   S1/2011\\
4006.4000   &   0.474   &   78    &   S3/2006 &  4675.5124   &   0.465   &   95    &   S3/2008  &  5726.6656   &   0.438   &   95    &   S1/2011\\
4007.3988   &   0.515   &   71    &   S3/2006 &  4676.5012   &   0.506   &   122    &   S3/2008  &  5727.6438   &   0.479   &   112    &   S1/2011\\
4008.3963   &   0.557   &   49    &   S3/2006 &  4677.5089   &   0.548   &   109    &   S3/2008  &  5730.6297   &   0.603   &   141    &   S1/2011\\
4008.6148   &   0.566   &   48    &   S3/2006 &   4678.4906   &   0.589   &   97    &   S3/2008  &  5731.6411   &   0.646   &   93    &   S1/2011\\
4009.3968   &   0.599   &   75    &   S3/2006 &  4679.5136   &   0.632   &   95    &   S3/2008  &  5733.6768   &   0.731   &   107    &   S1/2011\\
4010.3972   &   0.641   &   70    &   S3/2006 &  4680.4960   &   0.673   &   98    &   S3/2008  &  5737.6814   &   0.898   &   133    &   S1/2011\\
4011.3954   &   0.683   &   70    &   S3/2006 &  4681.4788   &   0.714   &   105    &   S3/2008  &  5738.6419   &   0.939   &   99    &   S1/2011\\
4012.4237   &   0.726   &   52    &   S3/2006 &  4682.4761   &   0.756   &   114    &   S3/2008  & 5739.6380   &   0.980   &   73    &   S1/2011\\
4013.3943   &   0.766   &   56    &   S3/2006 &  4683.4900   &   0.798   &   110    &   S3/2008  &  5740.6246   &   0.022   &   115    &   S1/2011\\
4014.4257   &   0.809   &   35    &   S3/2006 &  4684.5041   &   0.841   &   105    &   S3/2008  &  5741.6552   &   0.065   &   95    &   S2/2011\\
4014.5202   &   0.813   &   35    &   S3/2006 &  4685.4908   &   0.882   &   85    &   S3/2008  &  5742.6304   &   0.106   &   112    &   S2/2011\\
4015.6057   &   0.859   &   67    &   S3/2006 &  4687.4826   &   0.965   &   95    &   S3/2008  &  5743.6332   &   0.147   &   84    &   S2/2011\\
4017.3877   &   0.933   &   68    &   S3/2006 &  4688.4611   &   0.006   &   111    &   S3/2008  &  5744.6525   &   0.190   &   73    &   S2/2011\\
4018.4017   &   0.976   &   64    &   S3/2006 &  4689.4824   &   0.049   &   118    &   S3/2008  &  5745.6220   &   0.231   &   107    &   S2/2011\\
4019.3718   &   0.016   &   33    &   S3/2006 &  4692.4841   &   0.175   &   98    &   S3/2008  &  5746.6009   &   0.272   &   108    &   S2/2011\\
4019.4138   &   0.018   &   47    &   S3/2006 &  4693.4718   &   0.216   &   75    &   S3/2008  &  5749.5862   &   0.397   &   108    &   S2/2011\\
4020.3879   &   0.059   &   73    &   S3/2006 &  4694.4601   &   0.257   &   103    &   S3/2008  & 5751.6024   &   0.481   &   147    &   S2/2011\\
4021.3878   &   0.101   &   82    &   S3/2006 &  4695.4523   &   0.299   &   117    &   S3/2008  &   5752.5473   &   0.520   &   134    &   S2/2011\\
 4022.3871   &   0.142   &   61    &   S3/2006 &  4696.4539   &   0.341   &   94    &   S3/2008  & 5757.6309   &   0.733   &   134    &   S2/2011\\
  4618.6745   &   0.087   &   130    &   S1/2008 & 5020.6934   &   0.905   &   123    &   S1/2009  &  5758.5582   &   0.772   &   92    &   S2/2011\\
4619.6683   &   0.129   &   98    &   S1/2008 & 5021.6844   &   0.946   &   145    &   S1/2009  & 5759.5616   &   0.814   &   85    &   S2/2011\\
4620.6966   &   0.172   &   127    &   S1/2008 & 5022.6944   &   0.988   &   103    &   S1/2009  & 5760.5258   &   0.854   &   127    &   S2/2011\\
4621.6611   &   0.212   &   114    &   S1/2008 & 5023.6911   &   0.030   &   118    &   S1/2009  &                         &                  &               &                  \\
4622.6505   &   0.253   &   136    &   S1/2008 &  5027.5709   &   0.192   &   135    &   S1/2009   &                        &                  &               &                  \\
\hline
\end{tabular}
\end{footnotesize}
\tablefoot{
\tablefoottext{a} {2\,450\,000+}
\tablefoottext{b} {Phases computed using Eq.~\ref{eq1}.}
}
\end{table*}

% ---------------------------- FA1

\begin{figure*}[tb]
\centering
\includegraphics[width=0.575\columnwidth]{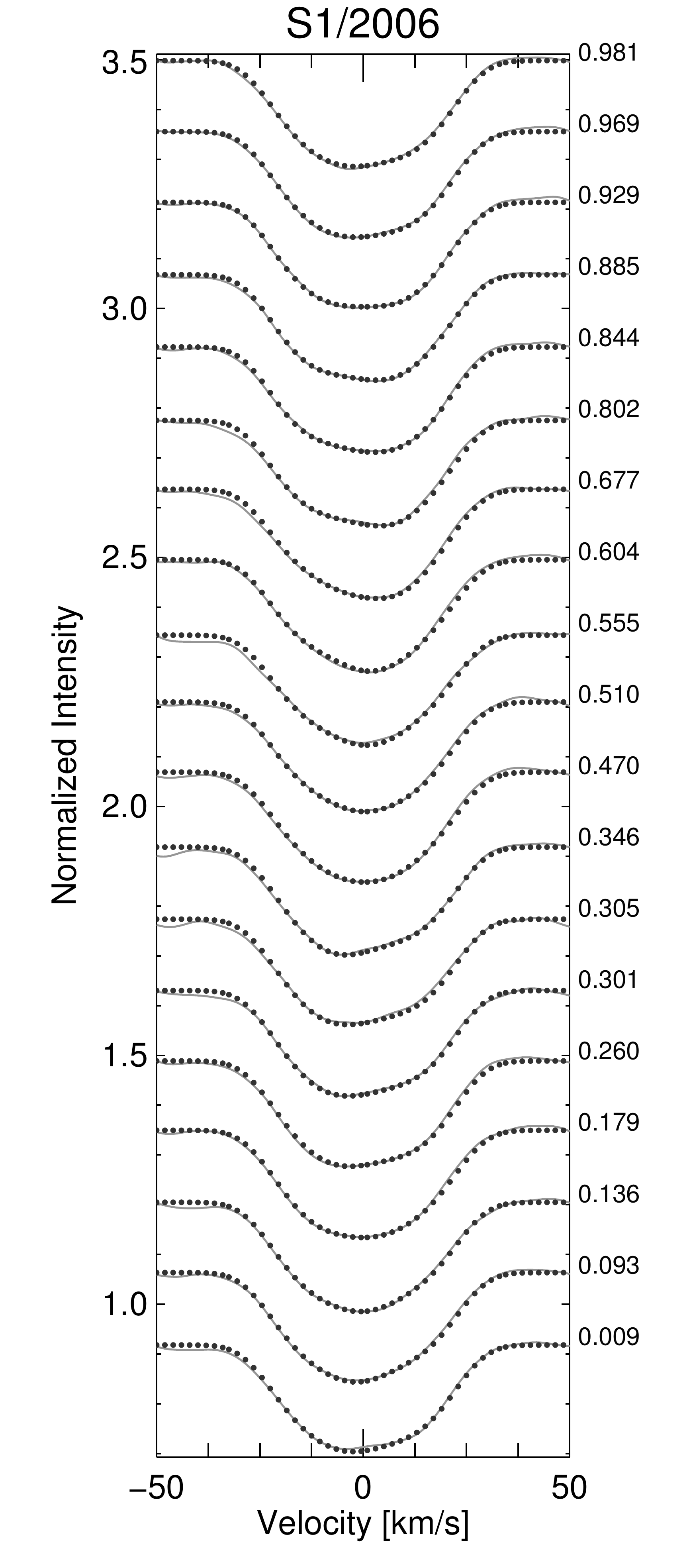}\hspace{9mm}
\includegraphics[width=0.575\columnwidth]{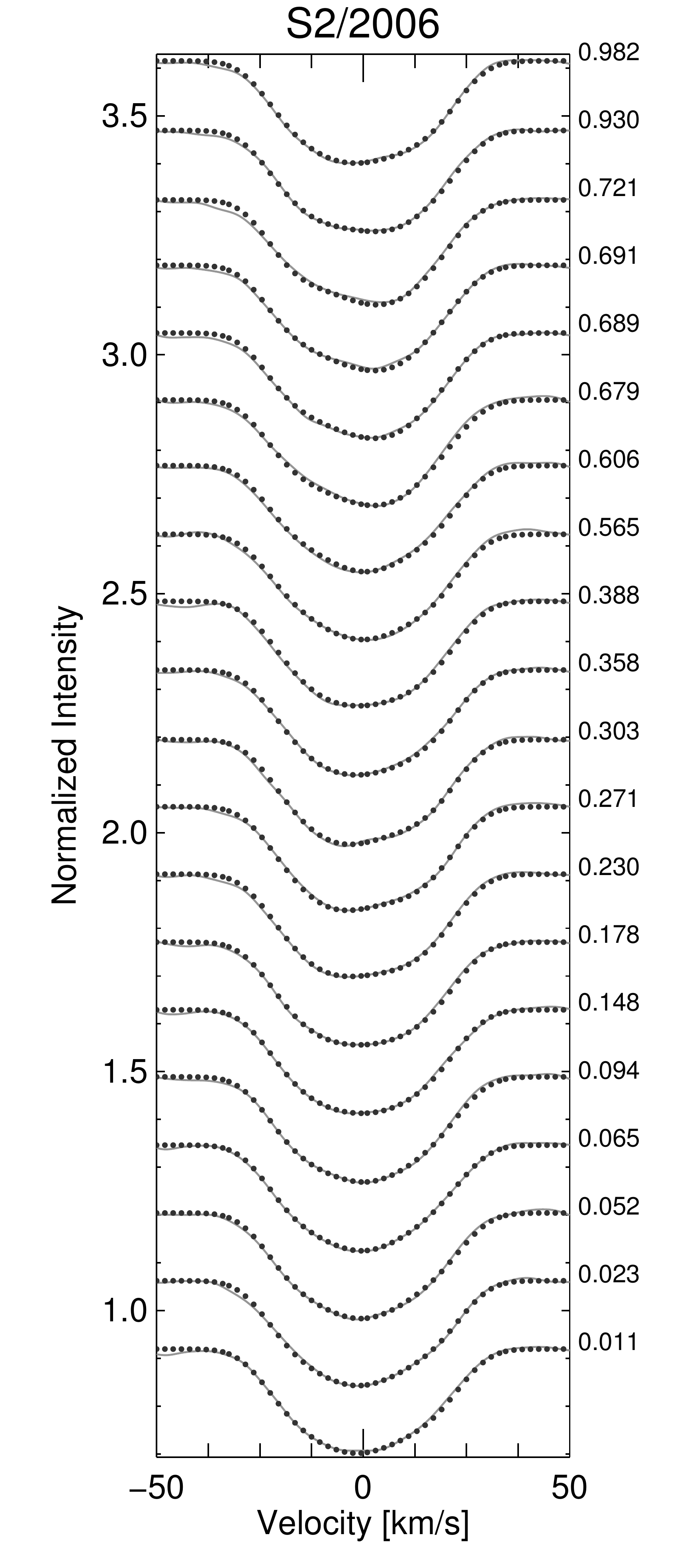}\hspace{9mm}
\includegraphics[width=0.575\columnwidth]{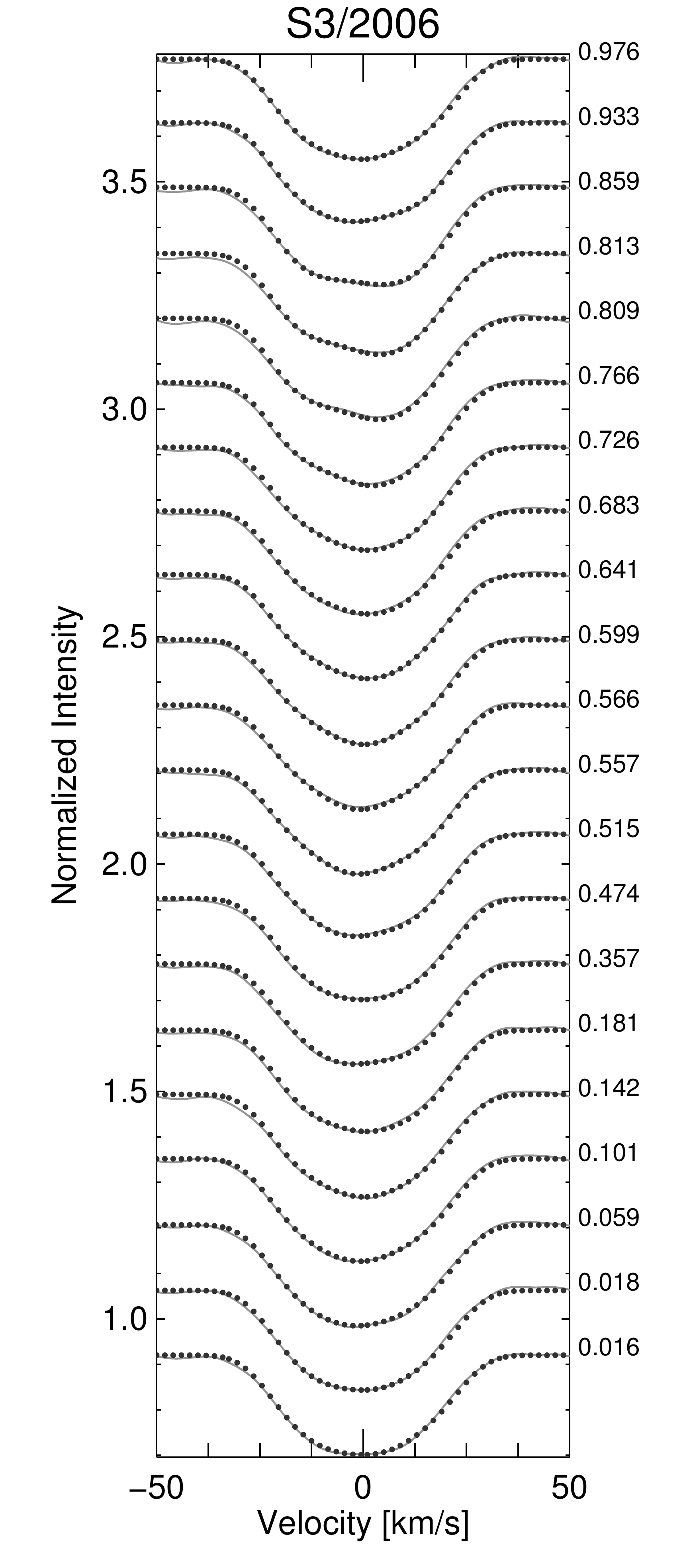}

\centering
\includegraphics[width=0.575\columnwidth]{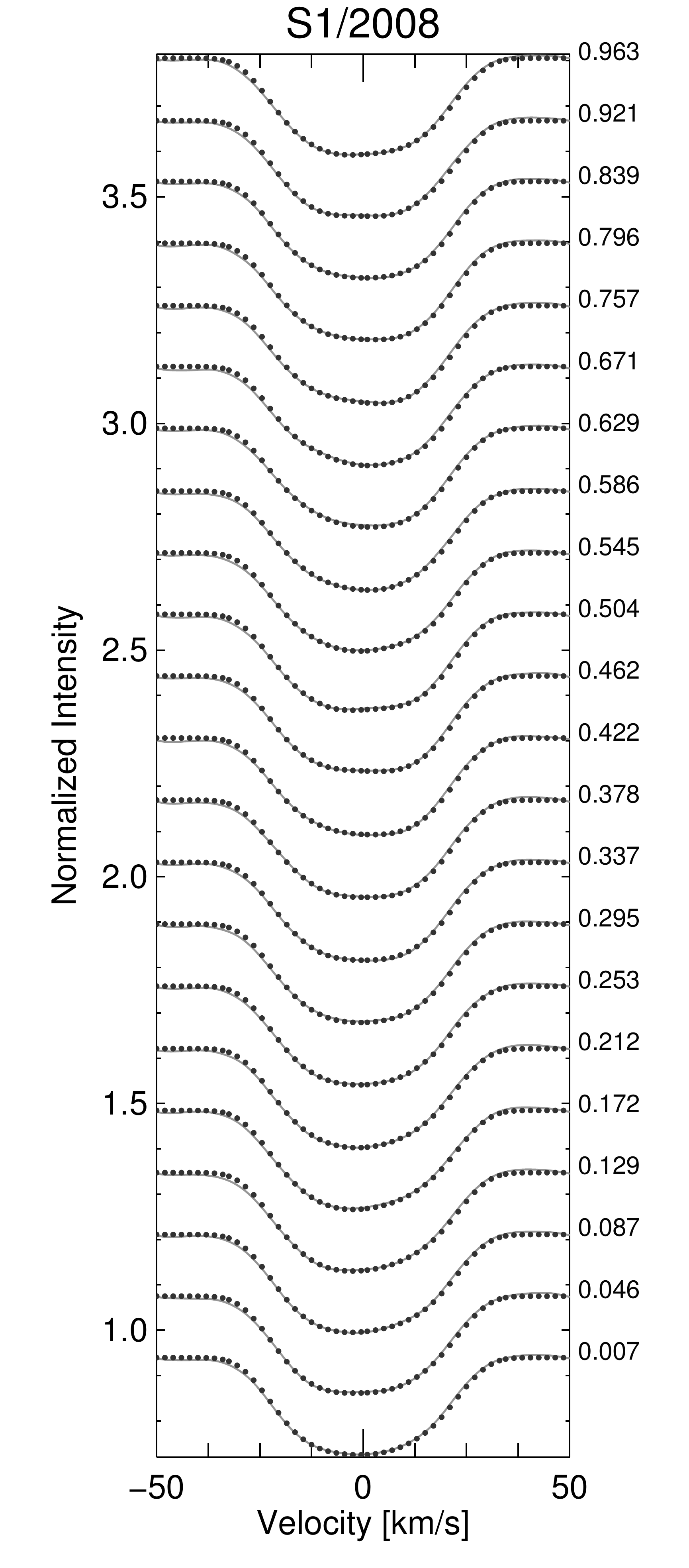}\hspace{9mm}
\includegraphics[width=0.575\columnwidth]{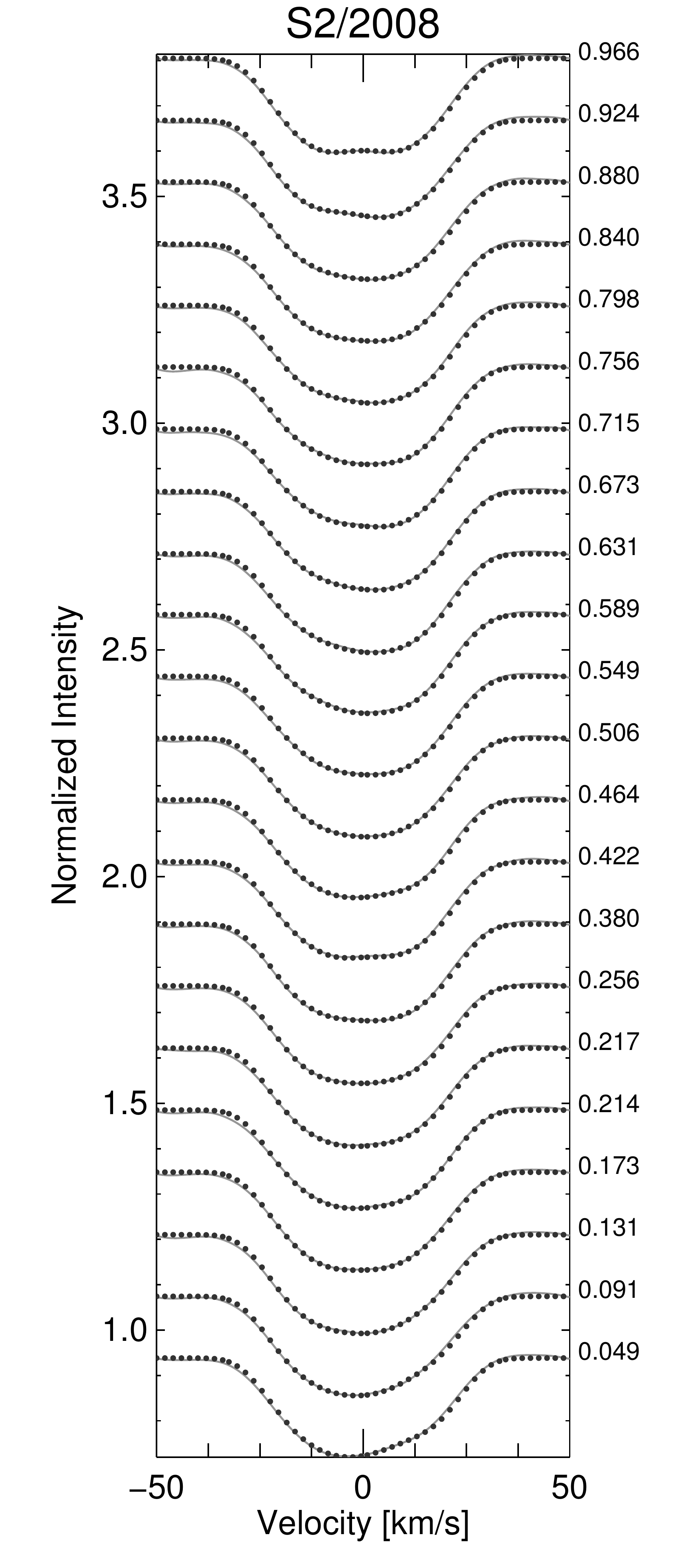}\hspace{9mm}
\includegraphics[width=0.575\columnwidth]{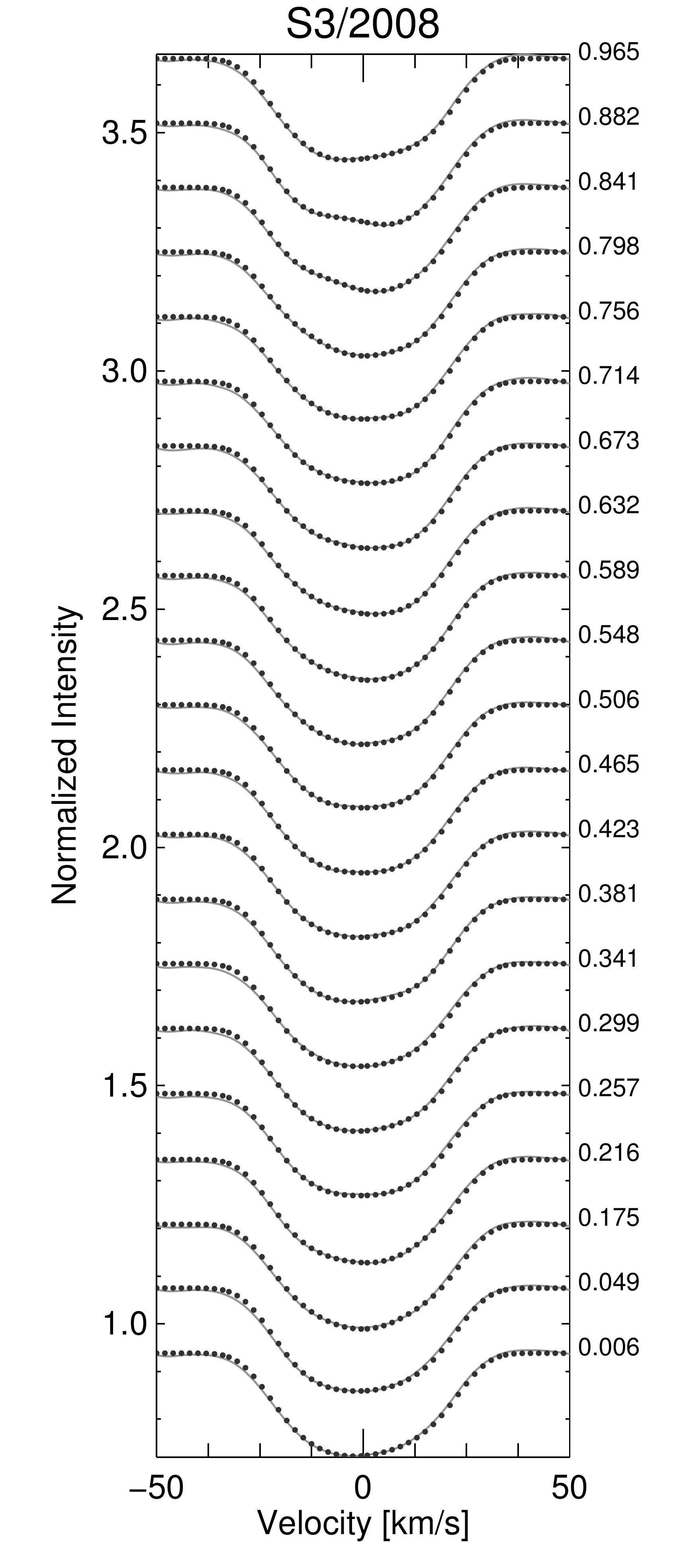}
\caption{Line profile fits for the Doppler reconstructions shown in Figs.~\ref{2006di}--\ref{2008di}.
The phases of the individual observations are listed on the right side of the panels.}
\label{proffits1}
\end{figure*}

% ---------------------------- FA2

\begin{figure*}[tb]
\centering
\includegraphics[width=0.575\columnwidth]{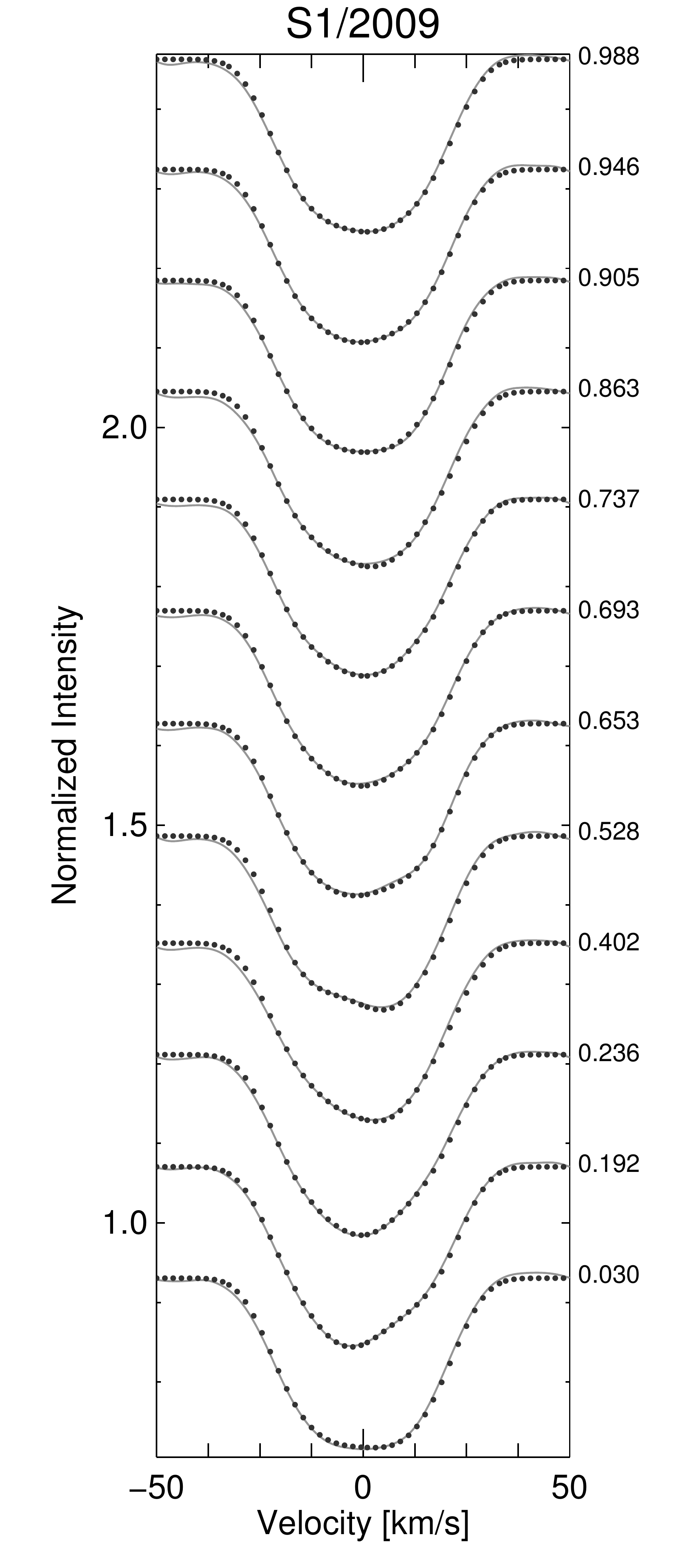}\hspace{9mm}
\includegraphics[width=0.575\columnwidth]{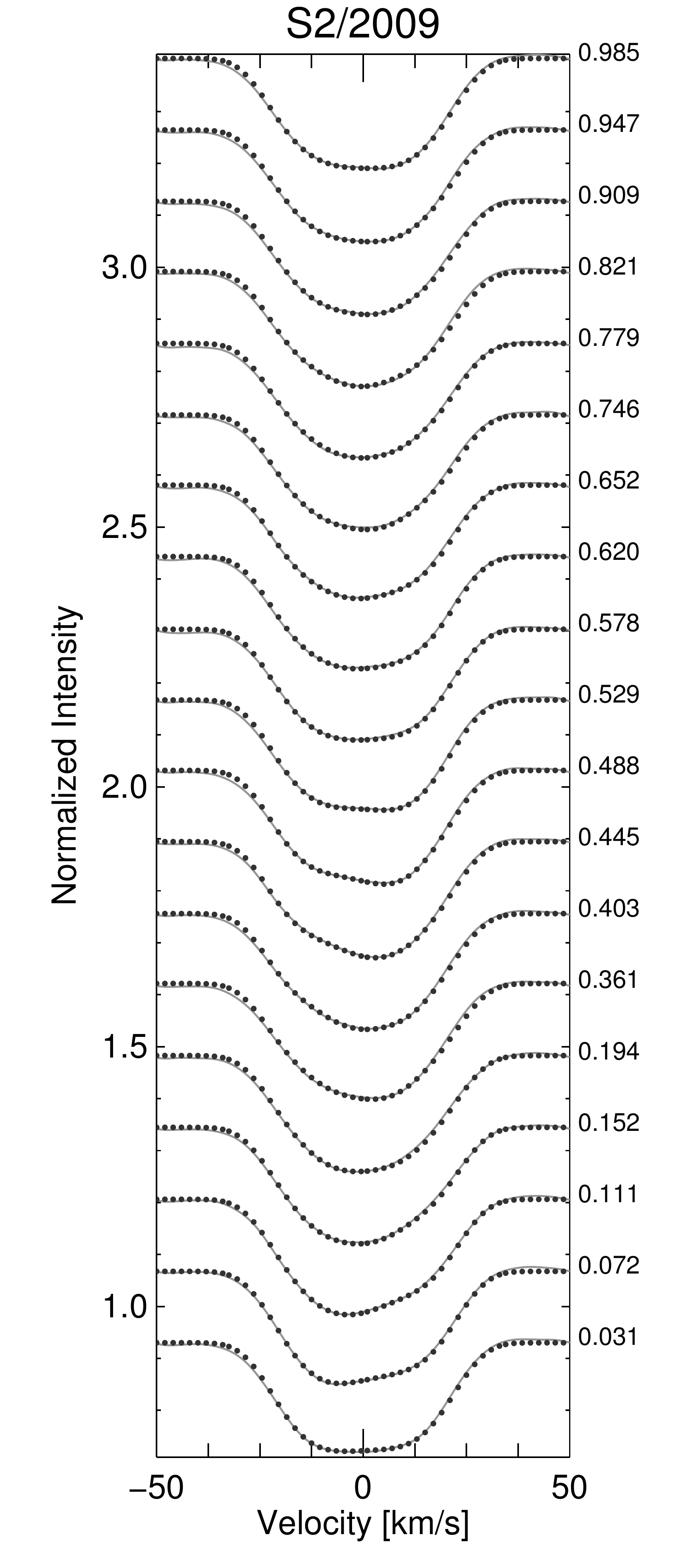}\hspace{9mm}
\includegraphics[width=0.575\columnwidth]{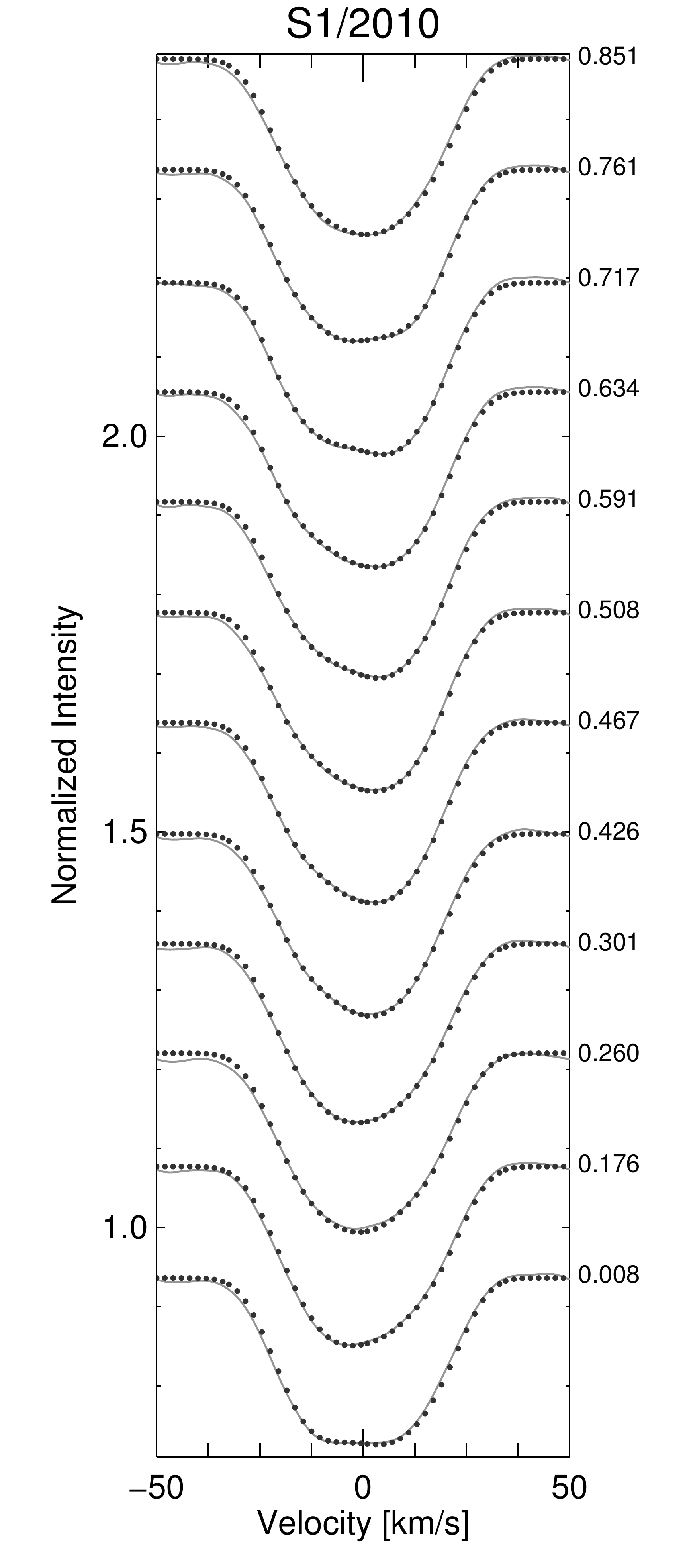}

\centering
\includegraphics[width=0.575\columnwidth]{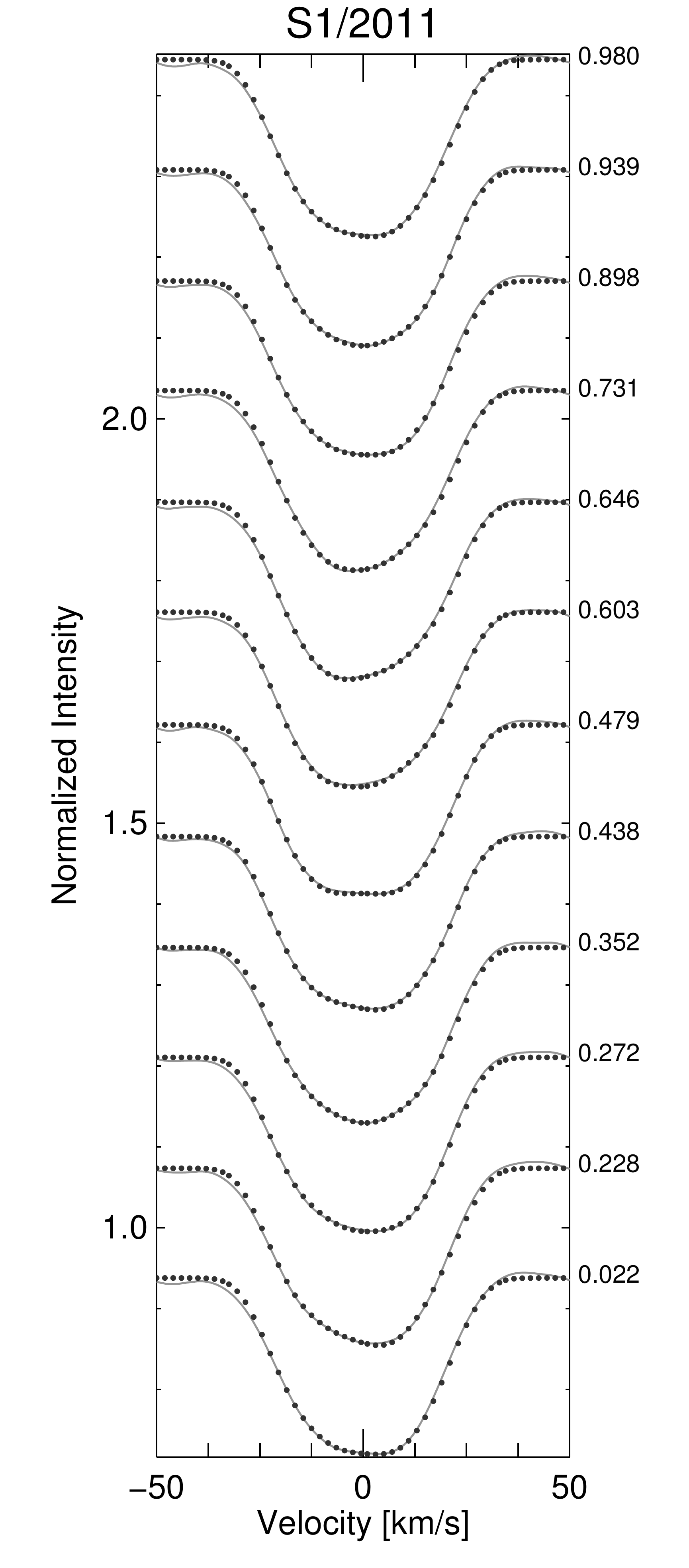}\hspace{9mm}
\includegraphics[width=0.575\columnwidth]{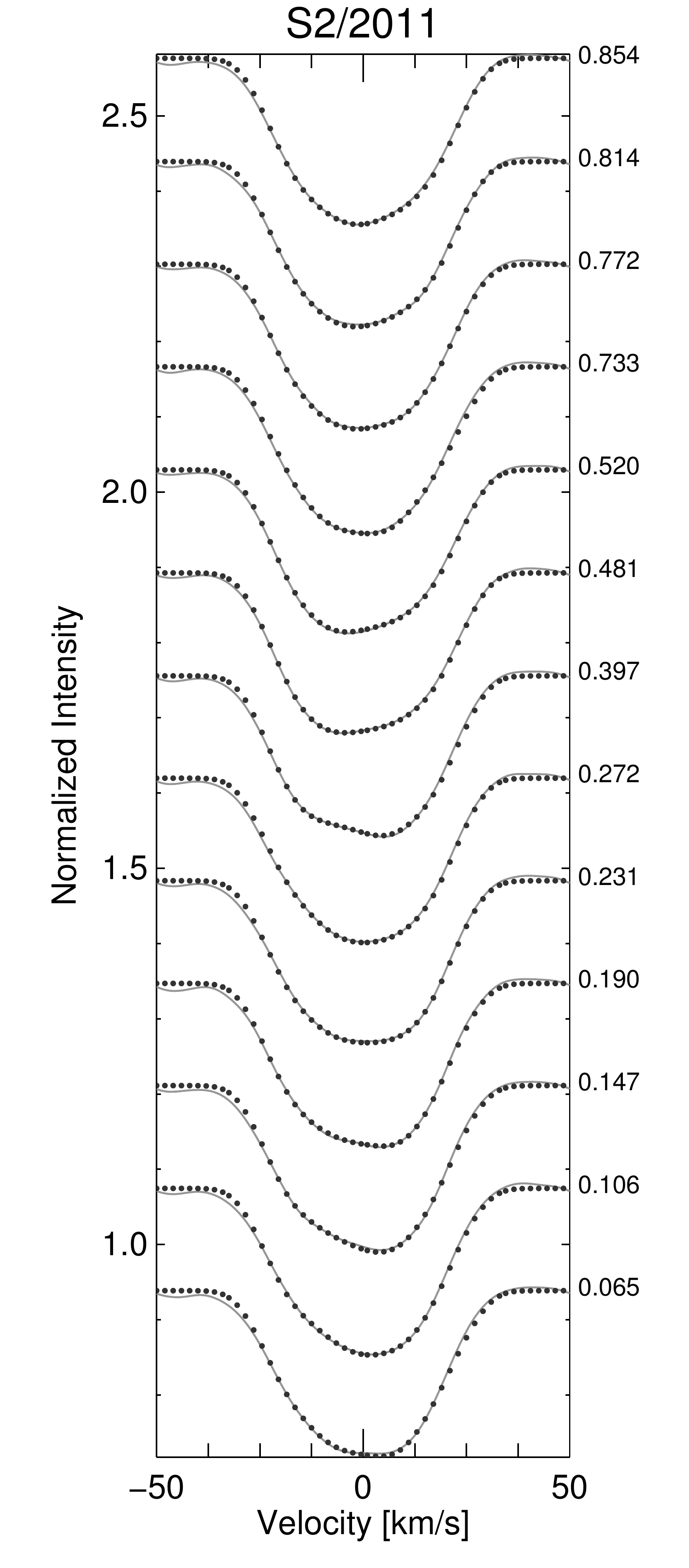}
\caption{Line profile fits for the Doppler reconstructions shown in Figs.~\ref{2009di}--\ref{2011di}.
The phases of the individual observations are listed on the right side of the panels.}
\label{proffits2}
\end{figure*}

\end{appendix}

\appendix
\end{document}